\documentclass{vgtc}                          
\pdfoutput=1
\let\ifpdf\relax

\usepackage{mathptmx}
\usepackage[pdftex]{graphicx}
\usepackage{times}

\usepackage{makeidx} 
\usepackage{cite}
\usepackage{latexsym}
\usepackage{amsmath}
\usepackage{amsthm}
\usepackage{amssymb}
\usepackage{graphicx}
\usepackage{verbatim}
\usepackage{url}
\usepackage{times}
\usepackage{compress}
\usepackage{subfig}

\def\withcomments{
   \newcounter{mycommentcounter}
   \def\aside##1{\refstepcounter{mycommentcounter}%
    \ifhmode%
     \unskip%
     {\dimen1=\baselineskip \divide\dimen1 by 2 %
       \raise\dimen1\llap{\tiny -\themycommentcounter-}}\fi%
     \marginpar{\renewcommand{\baselinestretch}{0.8}%
       \footnotesize\tiny [\themycommentcounter]: \raggedright ##1}}
   }
\withcomments

\onlineid{241}

\vgtccategory{System}

\vgtcinsertpkg

\title{Maps of Computer Science}

\author{Daniel Fried\thanks{e-mail: dfried@cs.arizona.edu}\\ %
        \parbox{2in}{\scriptsize \centering Department of Computer Science \\ University of Arizona, Tucson, AZ, USA}%
\and Stephen G.~Kobourov\thanks{e-mail: kobourov@cs.arizona.edu}\\ %
     \parbox{2in}{\scriptsize \centering Department of Computer Science \\ University of Arizona, Tucson, AZ, USA} %
}

\teaser{
    \vspace{-.1cm}\includegraphics[width=4in]{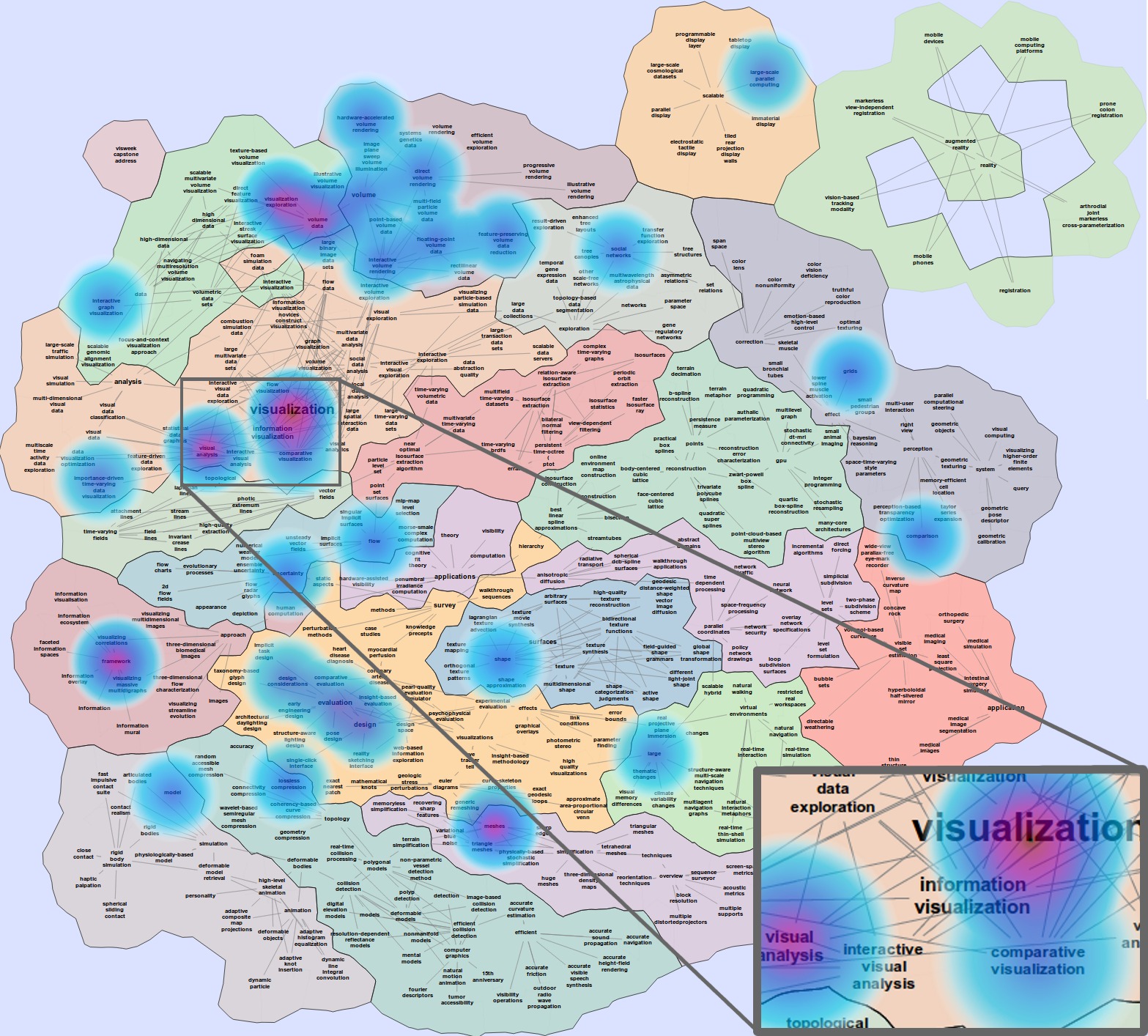}
  \caption{\normalsize Map of TVCG based on 1,343 TVCG titles in DBLP,
  heatmap overlay based on 34 papers by the most prolific TVCG author.
  (Multi-Word Term extraction, C-Value with Unigrams ranking, Partial Match Jaccard Coefficient similarity, Pull Lesser Terms
  filtering, number of terms 1500.)}
  \label{tvcg_kwan}
}

\abstract{  
We describe a practical approach for visual exploration of research papers. 
Specifically, we use the titles of papers
    from the DBLP database to create what we call {\em maps of computer
    science} ({\tt MoCS}).
Words and phrases from the paper titles are the cities in the map, and countries are created based on word and phrase similarity, calculated using co-occurence. 
 With the help of heatmaps, we can visualize the {\em
    profile} of a particular conference or journal over the base map.
    Similarly, heatmap profiles can be made of individual researchers or groups
    such as a department. The visualization system also makes it possible to
    change the data used to generate the base map. For example, a specific
    journal or conference can be used to generate the base map and then the
    heatmap overlays can be used to show the evolution of research topics in
    the field over the years. As before, individual researchers or research
    groups profiles can be visualized using heatmap overlays but this time over
    the journal or conference base map. Finally, research papers or abstracts
    easily generate {\em visual abstracts} giving a visual representation of
    the distribution of topics in the paper. We outline a modular and extensible
    system for term extraction using natural language processing techniques,
    and show the applicability of methods of information retrieval to
    calculation of term similarity and creation of a topic map. The system is
    available at \url{mocs.cs.arizona.edu}.
}

\begin{document}

\firstsection{Introduction}
\maketitle

Providing efficient and effective data visualization is a difficult
challenge in many real-world software systems. One challenge lies in
developing algorithmically efficient methods to visualize large and
complex data sets. Another challenge is to develop effective
visualizations that make the underlying patterns and trends easy to
see. Even tougher is the challenge of providing interactive access,
analysis, and filtering. All of these tasks become even more difficult
with the size of the data sets arising in modern applications. In this
paper we describe {\em maps of computer science} ({\tt MoCS}), a
functional visualization system for a large relational 
data set, based on spatialization and map representations.

{\em Spatialization} is the process of assigning 2D or 3D coordinates to
abstract data points, ideally in such a way that the spatial mapping
has much of the characteristics of the original (higher dimensional)
space. Multi-dimensional scaling (MDS), principal component analysis
(PCA), and force-directed methods are among the
standard techniques that allow us to spatialize high-dimensional data.

\begin{figure*}[t]
\centerline{
    \includegraphics[width=0.85\textwidth]{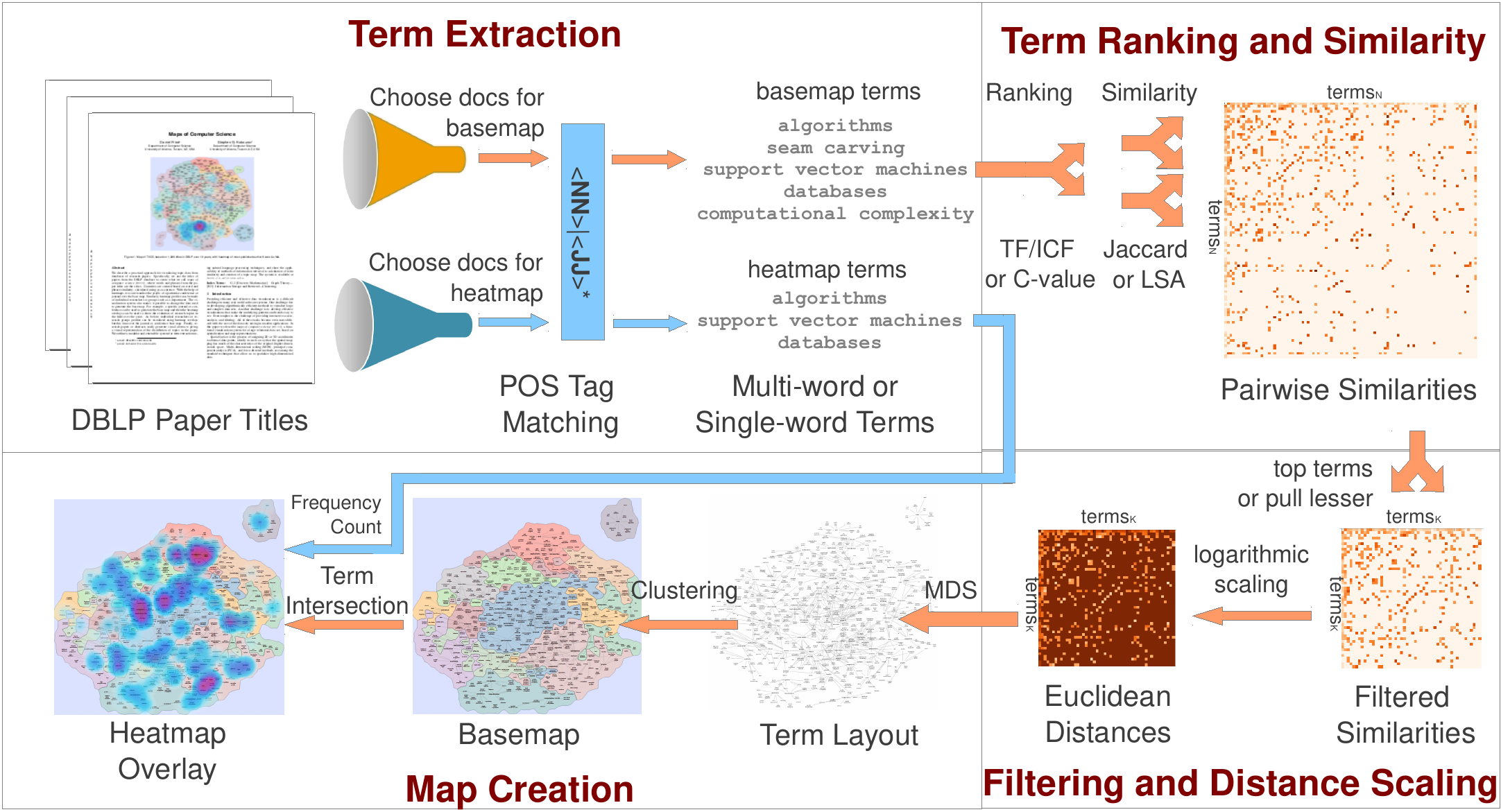}
}
    \caption{The main steps of the \texttt{MoCS} system are querying documents from
    DBLP, extracting terms from these titles, ranking terms by importance,
calculating term similarity, further filtering terms based on similarity, and finally
performing multidimensional scaling and clustering to produce a basemap, over which
a heatmap can be overlaid.}
    \label{system}
\end{figure*}
{\em Map representations} provide a way to visualize relational data with
the help of conceptual maps as a data representation metaphor. Graphs
are a standard way to visualize relational data, with the objects
defining vertices and the relationships defining edges.  It requires
an additional step to get from graphs to maps: clusters of
well-connected vertices form countries, and countries share borders
when neighboring clusters are tightly interconnected.

Traditional maps offer a natural way to present geographical data
(continents, countries, states) and additional properties defined with
the help of contours and overlays (topography, geology, rainfall).
In the process of data mining and data analysis, clustering is a very
important step. It turns out that maps are very helpful in dealing
with clustered data. There are several reasons why a map
representation of clusters can be helpful. First, by explicitly
defining the boundary of the clusters and coloring the regions, we
make the clustering information clear. Second, as most
dimensionality-reduction techniques lead to a two-dimensional
positioning of the data points, a map is a natural
generalization. Finally, while it often takes us considerable effort
to understand graphs, charts, and tables, a map representation is
intuitive, as most people are familiar with maps and map-based
interactions such as pan and zoom.

We describe a practical approach for visualizing data from
  the DBLP bibliography server~\cite{ley2009dblp}. Specifically, we use
  the titles of 2,184,270 papers in the
  database to create, what we call, {\em maps of computer science} ({\tt MoCS}),
  where words and phrases from the titles are the cities and where the
  countries are created based on co-occurrence. With the help of
  heatmap overlays, we can visualize the {\em profile} of a particular
  conference or journal over the base map. Similarly, individual
  researchers or groups such as a department can be used to generate
  heatmap profiles. The visualization system also makes it possible to
  change the data used to generate the base map. For example, a
  specific journal or conference can be used to generate the base map
  and then the heatmap overlays can be used to show the evolution of
  research topics in the field over the years with the help of small multiples. As before, individual
  researchers or research groups can profiles can be
  visualized using heatmap overlays but this time over the journal or
  conference base map. Finally, research papers or abstracts easily generate
  {\em visual abstracts}.

%
An overview of our {\tt MoCS} system is in Figure~\ref{system} and our main contributions are as follows. First,
we describe a fully functional visualization system {\tt MoCS} which 
interactively generates {\em base maps of computer science} from the DBLP
bibliography server: from maps based on all papers available in the
database, to maps based on a particular journal or conference, to maps
based on an individual researcher. Second, our system allows us to
visualize {\em temporal heatmap overlays} making it possible to
visualize the evolution of the field, journals, and conferences over
time. Third, our system allows us to visualize {\em individual heatmap
  overlays} making it possible to visualize individual researchers in
the field, or individual researchers in a particular conference, or 
individual papers in a particular conference. Finally, the {\tt MoCS} system is modular, extensible and with complete source code, thus making it easy to change various components: from the various natural language processing steps, to the creation of the graph that models the topics, to the visualization of the results.
%
%
%
%
%
%
%

\section{Related Work\label{sec_related}}

Using maps to visualize non-cartographic data has been considered in
the context of spatialization by Skupin and Fabrikant~\cite{sf-sm-03}
and Fabrikant {\em et al.}~\cite{Fabrikant_2006_map_infoviz}.  Map-like visualization
using layers and terrains to represent text document corpora dates
back at least to 1995 Wise {\em et al.} approach~\cite{857579}. %
Cortese {\em et
  al.}~\cite{Cortese_2006_contour} also use a topographical map metaphor to
visualize prefixes propagation in the Internet, where contour lines describing the propagation are calculated
using a force directed algorithm.  
The problem of effectively conveying change over time using a map-based visualization was studied by Harrower~\cite{harrower}.
 Also related is work on visualizing subsets of a set of items using
geometric regions to indicate the grouping. Byelas and
Telea~\cite{Byelas_2006_map} use deformed convex hulls to highlight
areas of interest in UML diagrams. Collins {\em et
al.}~\cite{Collins_2009_bubblset} use ``bubblesets,'' based on
isocontours, to depict multiple relations among a set of
objects. Simonetto {\em et al.}~\cite{Simonetto_2009_overlapset}
automatically generate Euler diagrams which provide one of the
standard ways, along with Venn diagrams, for visualizing subset
relationships. 

GMap uses the geographic map metaphor for visualizing
relational data and was proposed in the context of visualizing
recommendations, where the underlying data is TV shows and the
similarity between
them~\cite{Emden_Hu_recsys_2009,hu99visualizing}. This approach
combines graph layout and graph clustering, together with appropriate
coloring of the clusters and creating countries based on clusters and
connectivity in the original graph.  
A comprehensive overview of graph based representations by von
Landesberger {\em et al.}~\cite{CGF:CGF1898} considers visual graph
representation, interaction, editing, and algorithmic 
analysis.

Word clouds and tag clouds have been in use for many years~\cite{viegas-tag-08,clouds-07}. The
popular tool, Wordle~\cite{wordle} took word clouds to the next level with
high quality design, graphics, style and functionality. While these
early approaches do not explicitly use semantic information such as
word relatedness in placing the words in the cloud, several more
recent approaches do. Koh {\em et al.}~\cite{maniwordle} use interaction to add semantic
relationship in their ManiWordle approach. Parallel tag
clouds by Collins {\em et al.}~\cite{collins-09} are used to visualize evolution over time with the help of
parallel coordinates.
Cui {\em et al.}~\cite{Cui_2010_wordcloud} couple trend charts with
word clouds to keep semantic relationships, while visualizing evolution over time with help of force-directed
methods. 
Wu {\em et al.}~\cite{wu2011semantic} introduce a method for creating semantic-preserving word 
clouds based on a seam-carving image processing method and an application of bubble sets.
Paulovich {\em et al.}~\cite{CGF:CGF3107} combine semantic proximity with techniques for
fitting word clouds inside general polygons.
They apply
this technique to a collection of documents and obtain several word
clouds of related terms, while optimizing word packing into polygons 
with semantic preservation. Hierarchically clustered document collections have been the domain of many visualizations based on self-organizing maps~\cite{HKK96}, Voronoi diagrams~\cite{infosky}, and Voronoi treemaps~\cite{brandes12}. Of course, classical treemaps~\cite{treemap} and their variants are also often used to visualize text collections.

There is a great deal of related work on natural language processing, text
summarization, topic extraction and associated visualizations.  Statistical
topic modeling relies on machine learning techniques to extract semantic or
thematic topics from a text collection, e.g., via Latent Semantic
Analysis~\cite{deerwester1990indexing},
or Latent Dirichlet Allocation~\cite{Blei03}. Extensions to these topic models
allow discovery of topics underlying multi-word phrases~\cite{wang2007topical}
and the use of additional syntactic structure, such as sentence parse trees, to
aid inference of topics~\cite{boyd2010syntactic}.  The topics provide an
abstract representation of the text collection and are used for searching and
categorization. For example, Grouper~\cite{zamir1999grouper} presents search
results as sets of documents clustered by common phrases.
TopicNets~\cite{topicnets} assigns the top two words as a summary of the
underlying text.  TopicIslands~\cite{topicislands} is one of the early
visualizations, based on wavelets. More recently, Facetatlas~\cite{facetatlas}
uses similarity between documents to create a graph which can be used to
visually explore the data.  PhraseNets supports search for user provided
word-pairs which are then used to create graph-based visualization of
text~\cite{phrasenet}. TagRiver~\cite{forbesinteractive} uses word clouds to
visualize temporal changes in semantic data.  The TIARA system~\cite{tiara}
uses text summarization techniques and ThemeRiver-style
visualization~\cite{themeriver} to summarize large text collections.

\section{Maps of Computer Science}
Here we describe the main steps in the system: natural language processing (term extraction, term ranking, term filtering, similarity matrix), graph and map generation (distance matrix, embedding, clustering, coloring).
\subsection{Term Extraction}

\begin{figure}
    \includegraphics[width=.5\textwidth]{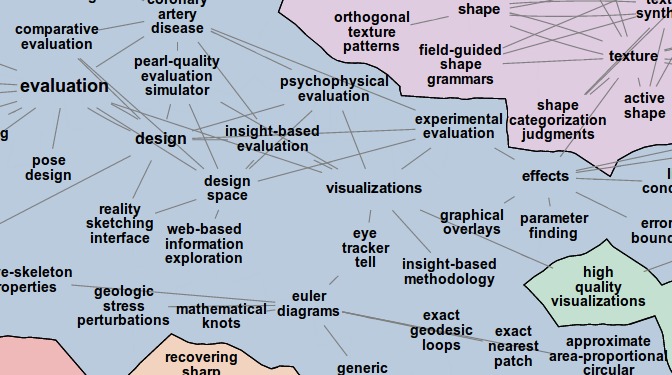}
    \caption{Section of a multi-word term map, built from 1,343 TVCG paper
    titles using the C-Value with Unigrams ranking, Partial Match Jaccard
Coefficient similarity, and Pull Lesser Terms filtering functions, with the
number of terms parameter set to 1500.}
    \label{tvcgmulti}
\end{figure}

\begin{figure}
    \includegraphics[width=0.5\textwidth]{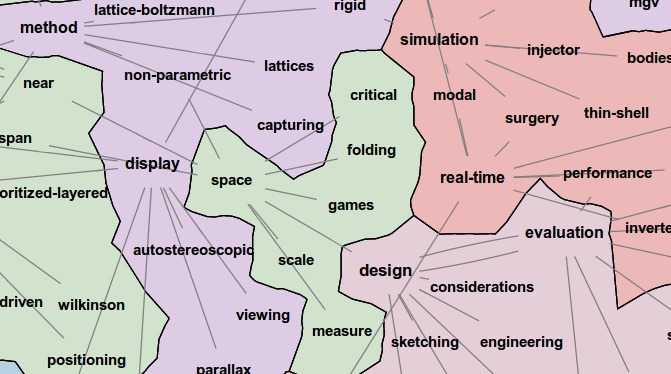}
    \caption{Section of a single-word term map, built from 1,343 TVCG paper
    titles using the TF ranking, LSA similarity, and Pull Lesser Terms
filtering functions, with the number of terms parameter set to 1800.}
    \label{tvcgsingle}
\end{figure}

In the first step of map creation, multi-word terms are extracted from the titles of
papers in DBLP. Part of speech (POS) tags are used to choose words that constitute
topically meaningful terms, and exclude functional words (words that
convey little semantic meaning, such as ``the'', ``and'', and ``a''). The Natural
Language Toolkit (NLTK) POS tagger~\cite{bird2009natural} is used to label the
words in all titles with POS tags.  Before running the tagger, titles are
converted to lowercase, since the tagger is case-sensitive, and more likely to
incorrectly label capitalized words as proper nouns.  Once a title is tagged,
maximal subsequences of words with POS tags matching the following regular
expression are extracted from titles:
\begin{equation*}
(\langle JJ\rangle |\langle JJR\rangle |\langle JJS\rangle |\langle NN\rangle |\langle NNS\rangle |\langle NNP\rangle |\langle NNPS\rangle )^*
\end{equation*}
$JJ, JJR,$ and $JJS$ are tags representing normal adjectives, comparative
adjectives, and superlative adjectives, respectively, while $NN$, $NNS$, $NNP$,
and $NNPS$ are nouns, plural nouns, proper nouns, and proper plural nouns,
respectively. This regular expression was chosen to extract a subset of noun
and adjectival phrases including modifiers such as noun adjuncts and
attributive adjectives.  For example, the paper title ``Interactive Support for
Non-Programmers: The Relational and Network Approaches'' is assigned the tag
sequence $JJ\;NN\;IN\;NNS\;DT\;JJ\;CC\;NN\;NNS$.  The subsequences $JJ\;NN$,
$NNS$, $JJ$, and $NN\;NNS$ are matched, and their corresponding word sequences
``interactive support'', ``non-programmers'', ``relational'', and ``network
approaches'' are extracted as terms.  Maps can be created with these multi-word
terms (Fig.~\ref{tvcgmulti}), or the terms can be broken up into their
constituent words (Fig.~\ref{tvcgsingle}) to parallel the word-based visual
representations of systems such as Wordle~\cite{wordle}. 


Maps and visualizations made from single words can display broad associations
between words, as demonstrated in the semantic word clouds of Wu {\em et
al.}~\cite{wu2011semantic}. Multi-word terms can provide a fine-grained view of
the topics represented in the database of paper titles.  For example, using
single-word terms extracted from the titles of 40,000 randomly sampled DBLP
papers, and the Latent Semantic Analysis similarity function (described below),
the 5 most similar terms to ``network'' are ``neural'', ``wireless',
``sensor'', ``analysis'', and ``model''. This list of similar terms helps
reveal that there are different types of networks. Using multi-word terms and
the C-value With Unigrams ranking function (described below), we find that the
terms ``neural network'' and ``wireless sensor network'' appeared frequently in
titles, and are both ranked in the top 1500 terms from this document set,
helping to explain why ``neural'', ``wireless'', and ``sensor'' were highly
associated with ``network'' in the single-word version. We can use multi-word
terms and similarity to investigate what topics are closely related to each
specific type of network. Using the Jaccard similarity function (described
below), the 4 terms ranked as most similar to ``neural network'' are
``predictions'', ``genetic algorithm'', ``dynamics'', and ``combinatorial
optimization problem'', while the 4 most similar terms to ``wireless sensor
network'' are ``reinforcement'', ``mobile robot'', ``modeling'', and
``energy'', showing disparate applications and related topics for the two
different types of networks.

\subsection{Term Ranking}
Once multi-word or single word terms are extracted, they can be assigned
importance scores, or \emph{weights}, based on their usage in the corpus of
titles. Terms are then ordered by their weights to produce a ranking of terms,
of which the top terms can be selected for inclusion in the visual map
representation. We implement four such ranking functions in the {\tt MoCS}
system: Term Frequency, Term Frequency/Inverse Comparison Frequency, C-Value,
and C-Value with Unigrams.

Under the \emph{term frequency} ranking function, each term's weight is the number of
times it occurred within the corpus. Term frequency tends to highly weight
functional words such as determiners and conjunctions: words that appear
frequently but convey little meaning such as ``the'', ``a''. In our system,
many of these functional words are already excluded by the term extraction step,
if their POS tags do not match the noun and adjectival phrase extraction expression. 
However, we still want to provide the option to exclude common phrases that
convey little semantic meaning, such as ``introduction'' (which occurs 9th in a
list of multi-word terms ordered by frequency from a 1,000,000 title sample of DBLP,
occurring 618 times).
To accomplish this, a standard modification to term-frequency is term
frequency--inverse document frequency (TF/IDF), where a term's weight in a text collection 
is proportional to its frequency in the document and
inversely proportional to the number of other documents it appears in. 
In our domain, consisting of many short documents (titles), terms usually only
occur once in each document, so the inverse document frequency of a term is
almost always 1. Therefore, we further modify TF/IDF to this corpus by treating the entire collection of titles as a
single document, and counting the term's frequency in a reference corpus from a
different domain to use as the inverse weighting value.  We refer to the resulting
method as \emph{term frequency--inverse comparison frequency} (TF/ICF). A
term's weight under TF/ICF is the number of times the term appeared in the
corpus of documents (\emph{target corpus}), divided by the number of times that
term appeared in a disparate corpus of text from a different domain (the
\emph{comparison corpus}):
\begin{equation*}
weight(t) = \frac{Target(t)}{Comp(t)}
\end{equation*}
In the above equation, $t$ is a term, $Target(t)$ is the count of times that
$t$ appeared in the target corpus (DBLP titles) as a complete term, and
$Comp(t)$ is the count of times that $t$ appeared in the comparison corpus.
The {\tt MoCS} system currently uses the Brown Corpus~\cite{francis1979manual}, a selection
of English text drawn from newspapers, fiction, and other wide-distribution
literature, as the comparison corpus.

\emph{C-value}~\cite{frantzi2000automatic} is specifically designed for
multi-word term ranking, accounting for possible nesting of multi-word terms
(where short terms appear as word subsequences of longer terms).  C-value
incorporates total frequency of occurrence, frequency of occurrences of the
term within other longer terms, the number of types of these longer terms, and
the number of words in the term.  The weight assigned by C-value is
proportional to the logarithm of the number of words in a term, so we also
include a modified implementation, \emph{C-value With Unigrams}, that adds one
to this length before taking the logarithm. This modification allows single
word terms to be assigned non-zero weight and be included in the set of top
terms.

After terms are assigned importance weights, they are sorted in order of
descending weight, and the top $N$ terms are selected for possible inclusion in
the map. $N$ (Number of Terms) is a configurable parameter passed to the {\tt
MoCS} system. Larger values of $N$ produce maps that include terms ranked lower
by the chosen ranking algorithm, i.e., words with lower weighted term frequency
in the set of titles queried.

\subsection{Similarity Matrix Computation}
Once a set of top terms is selected, pairwise similarity values between top
terms are calculated.  We seek similarity functions that measure how closely
the topics represented by two terms are related. Terms that refer to the same
or similar topic, or topics that are closely associated, should receive high
similarity values. We use term-document co-occurrence as the basis of these similarity values, 
assuming that terms that appear together in multiple documents (paper titles)
are more likely to be related in meaning.

The similarity functions take a term-document matrix, $M$, as input.  The
columns of $M$ correspond to titles of papers from DBLP, and rows correspond to
terms extracted by the term-extraction step.  The entries in the matrix are
calculated as
\begin{equation*}
M_{ij} = occurrences_j(term_i)
\end{equation*}
where $occurrences_j(term_i)$ is the number of times the term indexed by $i$
appeared in the document indexed by $j$.  We implement three similarity functions in the {\tt MoCS} system: Latent Semantic Analysis, Jaccard Coefficient, and Partial Match Jaccard coefficient.

\emph{Latent Semantic Analysis (LSA)}, described by Deerwester {\em et al.}~\cite{deerwester1990indexing}, is a method of extracting underlying semantic
representation from the term-document matrix, $M$. A low-rank approximation to
the term-document matrix is used to calculate the distance between terms in a
vector-space representation reflecting meaning in topical space. The singular
value decomposition
\begin{equation*}
    M = U \Sigma V^\top
\end{equation*}
is calculated using sparse-matrix methods. Rows in the product $U\Sigma$
represent terms as feature vectors in the high-dimensional semantic space. Terms are compared
using cosine similarity~\cite{manning2008introduction} of the feature vectors to
produce a matrix of pairwise similarities between terms. The cosine similarity
of two term vectors $v_i, v_j$ is calculated as 
\begin{equation*}
    \cos(\theta) = \frac{v_i \cdot v_j}{||v_i||\;||v_j||}
\end{equation*}
The value returned by this function is bounded between 0, indicating a maximal
angle between the term vectors in semantic space and no similarity between the
terms, and 1, indicating the term vectors, measuring decomposed co-occurrence,
are identical.

LSA is a standard approach to calculating term and document similarity in
information retrieval. However, as in the term ranking stage, terms rarely
occur more than once in a single document (particularly if they are multi-word
terms). In our case, the entries in the term-document matrix are effectively
boolean. Depending on the term-ranking algorithm used to select the most
important terms, the term-document matrix can also be quite sparse.

We provide {\em Jaccard coefficient}~\cite{jaccard1901etude} as an alternative
similarity function to accommodate the nearly boolean nature of the term-document
matrix. Jaccard calculates pairwise term similarity as the number of documents two
terms appeared together in, divided by the number of documents either term
appeared in:
\begin{equation*}
    Jacc(S_i, S_j) = \frac{|S_i \cap S_j|}{|S_i \cup S_j|}
\end{equation*}
where $S_i$ and $S_j$ are the sets of documents that the two terms being
compared appeared in. Like LSA, Jaccard Coefficient produces a value between 0,
indicating terms did not appear together in any documents and have no
similarity, and 1, indicating terms never appeared separately, and have maximal
similarity.  Jaccard coefficient alone treats terms as atomic units: multi-word
terms only match if they are identical. This approach produces very sparse
similarity matrices when used with a ranking algorithm such as C-value that
prioritizes multi-word terms. 

\emph{Partial Match Jaccard Coefficient}, attempts to address the sparsity of the C-value matrices, by treating two terms as identical for
the purpose of co-occurrence calculation if they contain a common subsequence of words.
For example, if  ``partial match jaccard coefficient'' and ``similarity'' both
occurred as multi-word terms in a paper title, and ``similarity'' and ``jaccard
coefficient'' were present in our list of top-terms but ``partial match jaccard coefficient''
was not, this function would count a co-occurrence between ``similarity'' and
``jaccard coefficient'' because the top term is a subsequence of the longer
term found in the title.

\subsection{Term Filtering and Distance Calculation}
Term similarities have been calculated between the $N$ highest ranked terms in
the previous step.  The next stage in the pipeline is \emph{filtering},
choosing the terms to include in the map.  We implement two filtering methods
in the {\tt MoCS} system: Top Terms and Pull Lesser Terms.

\emph{Top Terms} is the simplest type of filtering, where we take the top-ranked $K$ terms from the $N$ highest ranked terms $(K \le N)$. %
The default for $K$ in our current system is 150.
In practice, sparsity of data causes this method to produce fragmented maps, as
the top $K$ terms often have low similarity to other top terms (particularly
when the multi-word term-extraction system is used).

\emph{Pull Lesser Terms} attempts to address the fragmentation of the top terms method, by
using not only the highest ranked terms, but also maps lesser-ranked terms if they are
similar to a top-ranked term. Specifically, this method takes as input the $N$
highest ranked terms, $terms_N$, and their pairwise similarities, as
calculated in the ranking and similarity steps of the pipeline. The method
plots the $K$ highest ranked terms, $terms_K$, from among $terms_N$, and the
$l$ most similar terms from $terms_N$ for each term in $terms_K$. These $l$
most similar terms are plotted regardless of whether they are members of the
set $terms_K$.  Effectively, this method pulls in terms beyond the top $K$, if
they are more similar to a top term than any of the other top terms. The
default parameter values for $K$ and $l$ in our current system are $K = 90$, $l
= 8$.

The pairwise term similarity matrix is next converted into a matrix of
distances for use by the multi-dimensional scaling or force-directed algorithms
of GMap. Let $S(t_i, t_j) \in [0,1]$ be the similarity between two terms, calculated using
either LSA, Jaccard Coefficient, or Partial Match Jaccard Coefficient. Some choices
of document sets and ranking and similarity functions produce terms with a
similarity distribution more narrow than the theoretical range of the similarity
function, so rescaled similarity values are calculated as
\begin{equation*}
    \hat{S}(t_i, t_j) = \frac{S(t_i,t_j)}{\max_{m, n: m \ne n}S(t_m, t_n)}.
\end{equation*}

The distance between these two terms, $D(t_i, t_j)$, is calculated using these rescaled
similarity values as
\begin{equation*}
    D(t_1, t_2) = -\log[(1 - \sigma)\cdot \hat{S}(t_1, t_2) + \sigma],
\end{equation*}
where $\sigma$ is a small, positive, constant scaling value, currently set to $0.1$,
used to ensure a non-zero value inside the logarithm in the case that two terms
have a pairwise similarity of 0. Linear transformations of similarities into distances produced maps
that looked dense, crowded, and highly fragmented.  A logarithmic scale allows comparison
of relative distance between terms with low pairwise similarity by magnifying
the distances between these terms. This produces more less crowded maps, since
most term pairs have low pairwise similarity compared to the highest similarity
pair of terms in the map (which are used in the normalization). 

\subsection{Map Generation}

We begin with a summary of the GMap algorithm for generating maps from static
graphs~\cite{hu99visualizing}. The input to the algorithm is a set of terms and
pairwise similarities between these terms, from which an undirected graph
$G=(V,E)$ is extracted. The set of vertices $V$ corresponds to the terms
extracted from titles and the set of edges $E$ corresponds to the top pairwise
similarities between these terms as determined by the chosen filtering
algorithm.  In its full generality, the graph is vertex-weighted and
edge-weighted, with vertex weights corresponding to some notion of the
importance of a vertex and edge weights corresponding to some notion of the
closeness between a pair of vertices. In the \texttt{MoCS} system, the relative
frequencies of terms are used to determine the font size of the node label, using
a linear scale with the minimum frequency term producing the smallest 
label and the maximum frequency term producing the largest label.
The weight of an edge can be defined by the strength of the similarity between
a pair of words or terms, and these edges can be marked in the base map of terms.

In the first step of GMap the graph is embedded in the plane using a scalable
force-directed algorithm~\cite{Fruchterman_Reingold_1991} or multidimensional
scaling (MDS)~\cite{mds}.  In the second step, a cluster analysis is performed
in order to group vertices into clusters, using a modularity-based clustering
algorithm~\cite{Newman_2006}.

We use information from the clustering to guide the MDS-based layout.  In the
third step of GMap, the geographic map corresponding to the data set is
created, based on a modified Voronoi diagram of the vertices, which in turn is
determined by the embedding and clustering. Here ``countries'' are created from
clusters, and ``continents'' and ``islands'' are created from groups of
neighboring countries. Borders between countries and at the periphery of
continents and islands are created in fractal-like fashion. Finally, colors are
assigned with the goal that no two adjacent countries have colors that are too
similar.  In the context of visualizing dynamic data where the relative change
of popularity of important, we also use a heatmap overlay to highlight the
``hot'' regions.  Further geographic components can be added to strengthen the
map metaphor.  For instance, edges can be made semi-transparent or even
modified to resemble road networks. In places where there are large empty
spaces between vertices in neighboring clusters, lakes, rivers, or mountains
can be added, in order to emphasize the separation.

\subsection{Heatmap Overlays}
To visualize the profile of a \emph{target query} set of papers (for example,
papers from a specified time range, author, conference, or journal) over a map,
we use heatmap overlays.  Heatmaps highlight the terms in the basemap that also
occur in the target query, with color intensity proportional to the frequency
of the term's occurrence in the heatmap query.  Separate database queries are
issued for the papers used to produce the basemap and heatmaps (Fig.~\ref{system}), allowing a subset of the papers chosen for the basemap to be
used as the target query.  For example, a basemap can be constructed from a
sample of all available papers, and a heatmap constructed from all papers for a
particular journal (Fig.~\ref{tvcg_heatmap}), or a heatmap of a single author
can be overlaid on a basemap of papers from a journal that author frequently
publishes in (Fig.~\ref{tvcg_kwan}).

Whichever type of terms are chosen for the basemap (multi-word or single-word)
are also used to construct the heatmap overlay. If $terms_B$ is the set of
terms found in the query used to produce the basemap, and $terms_T$ is the set
of terms from the documents to be visualized in the heatmap, the terms
highlighted in the heatmap are the intersection of these two groups, $terms_H =
terms_B \cap terms_T$. The heatmap intensity, $I(t)$, of each term $t$ in
$terms_H$ is the number of times $t$ appeared in documents in the target query.
These intensities are transformed on a logarithmic scale to allow terms with
low $I$ values to be visible in the heatmap, and then normalized so that the
most frequently appearing term has intensity 1. The final normalized and
rescaled intensity value,
$\hat{I}(t)$ is
\begin{equation*}
    \hat{I}(t) = \frac{\log (I(t) + \beta)}{\max_{\hat{t} \in terms_H} \log(I(\hat{t}) + \beta)}
\end{equation*}
where $\beta$ is a small additive constant (currently set to 1) that ensures
terms that only appeared once in the heatmap query still receive a positive
$\hat{I}(t)$ value.

Basemaps are rendered in the browser as vector graphics, and heatmaps are drawn
as a semi-transparent raster overlay using the OpenLayers heatmap
implementation. This implementation uses a radial gradient centered at terms
with a defined intensity value, where the color intensity at the center of the
term is proportional to the $\hat{I}$ value for the term. Currently, the radius
of diffusion for the radial gradient is constant across all terms in a map and
chosen to correspond to roughly half the average distance between terms. To
ensure that each term in $terms_H$ has an overlay that exactly covers its
visual area in the basemap, this method might be improved by making the
diffusion radius for each term a function of the distance to the closest term
in the map.  Alternatively, Inverse Distance Weighting could be used to
calculate color intensities for all points over the basemap based on the
heatmap intensity values of all terms.

\section{DBLP Visualization}
\subsection{Individual Heatmap Overlays}
\begin{figure}
    \includegraphics[width=\linewidth]{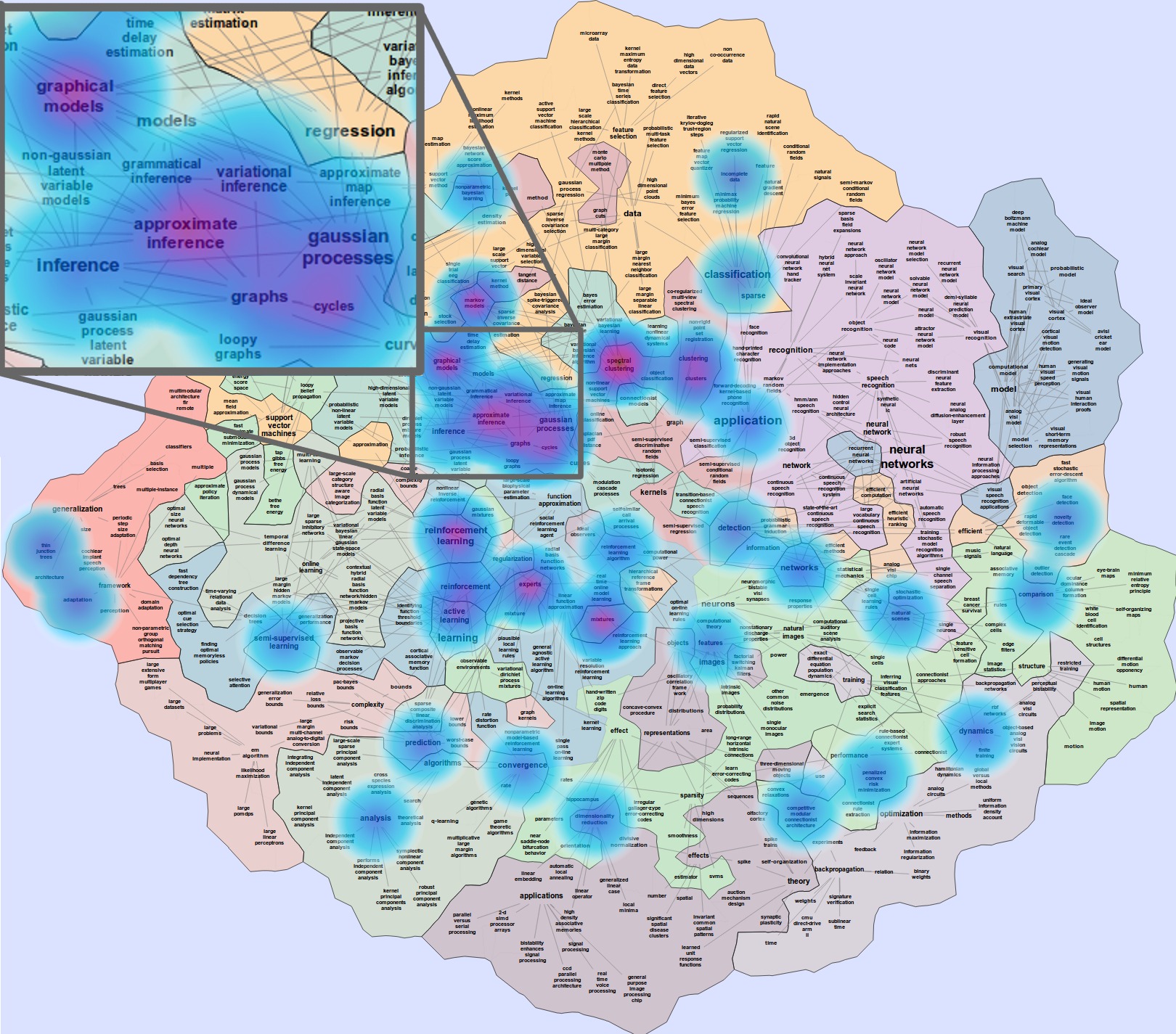}  
    \caption{A heatmap produced from 75 papers by the author who has published
        most frequently at NIPS, over a basemap made from multi-word terms
        extracted from titles of 3,553 NIPS papers. The algorithms used to
        produce the basemap are C-Value with Unigrams ranking, Partial Match
        Jaccard Coefficient similarity, and Pull Lesser Terms filtering, with the
        number of terms parameter set to 1,100.}
    \label{nips_jordan}
\end{figure}

The {\tt MoCS} system allows separate database queries for the documents used
to produce the basemap and the documents used to produce the heatmap overlay.
Using the author information in DBLP, we can produce heatmap overlays of
individual researchers over conferences and journals that they frequently
publish in. Figure~\ref{nips_jordan} shows a basemap constructed from titles of
all papers published at the Conference on Neural Information Processing Systems
(NIPS), with a heatmap constructed from the titles of papers by the most
prolific author at NIPS. We see activity throughout the basemap, with
particular intensity over a section of terms referring to inference in
graphical models.

\subsection{Conference and Journal Overlays}
\begin{figure*}[htb!]
    \centering
    \subfloat[Heatmap for CVPR made from 3,665 documents]{
        \includegraphics[width=0.45\linewidth]{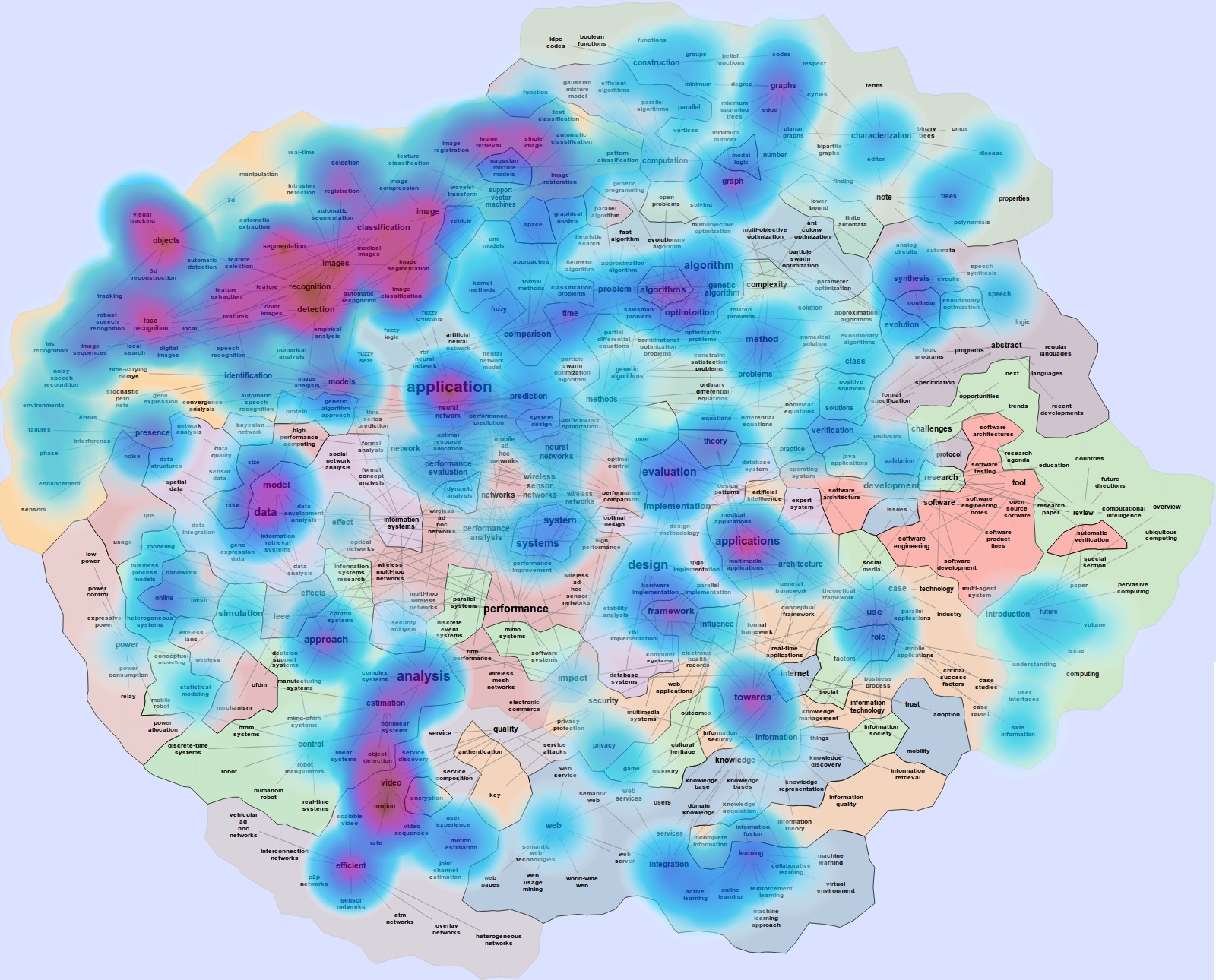}
        \label{cvpr_heatmap}
    }
    \subfloat[Heatmap for STOC made from 2,685 documents]{
        \includegraphics[width=0.45\linewidth]{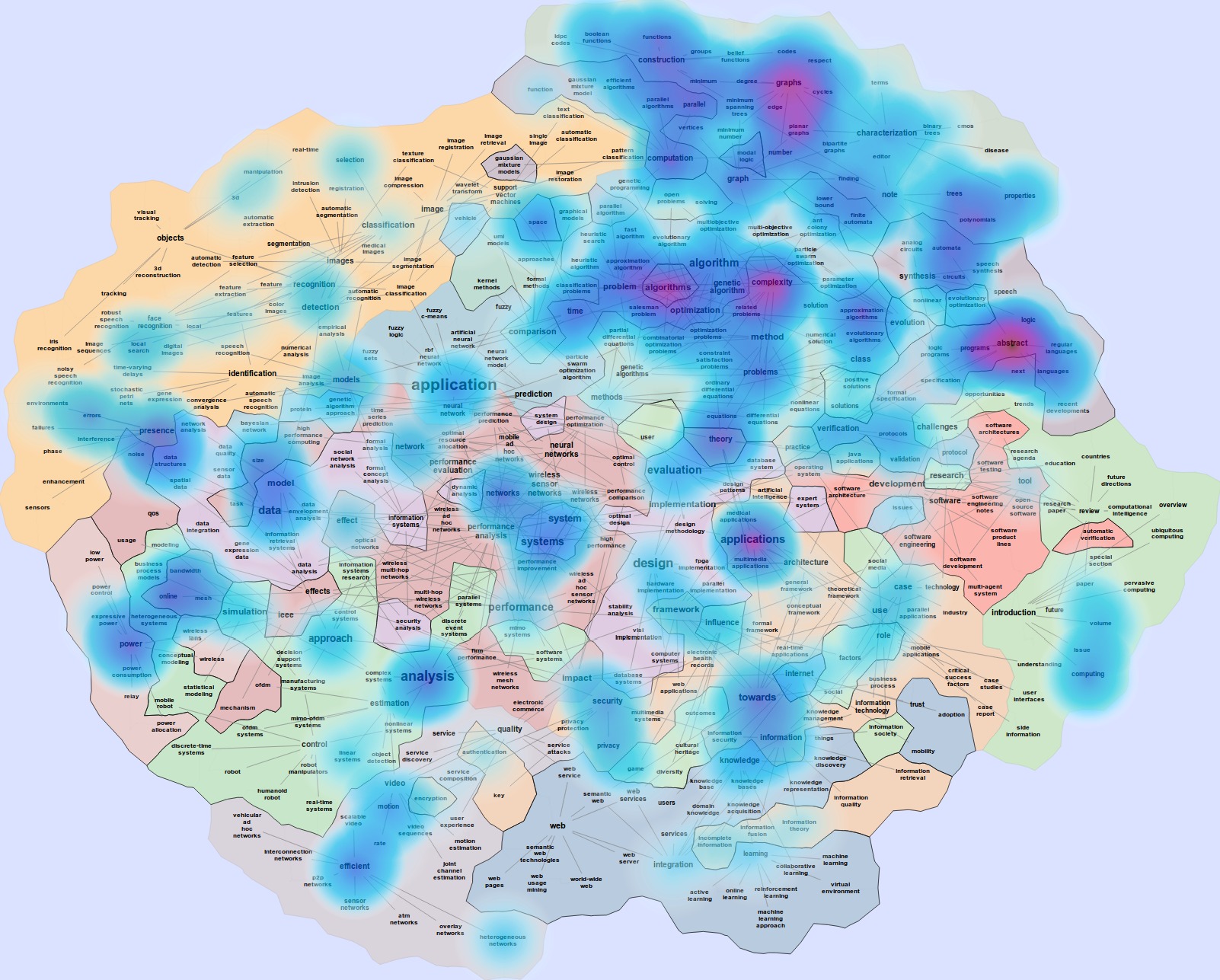}
        \label{ipl_heatmap}
    }

    \subfloat[Heatmap for ICWS made from 1,288 documents]{
        \includegraphics[width=0.45\linewidth]{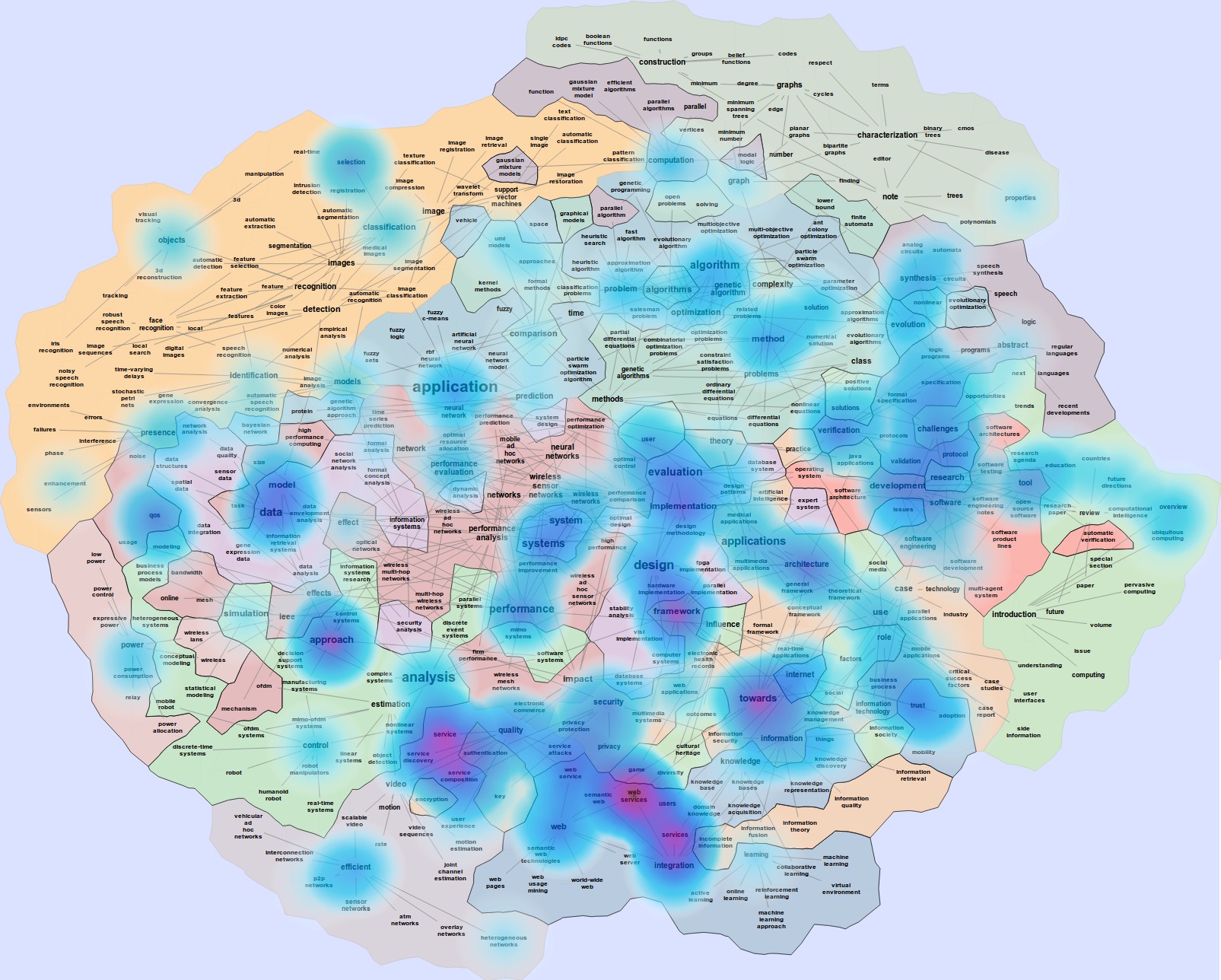}
        \label{icws_heatmap}
    }
    \subfloat[Heatmap for TVCG made from 1,826 documents]{
        \includegraphics[width=0.45\linewidth]{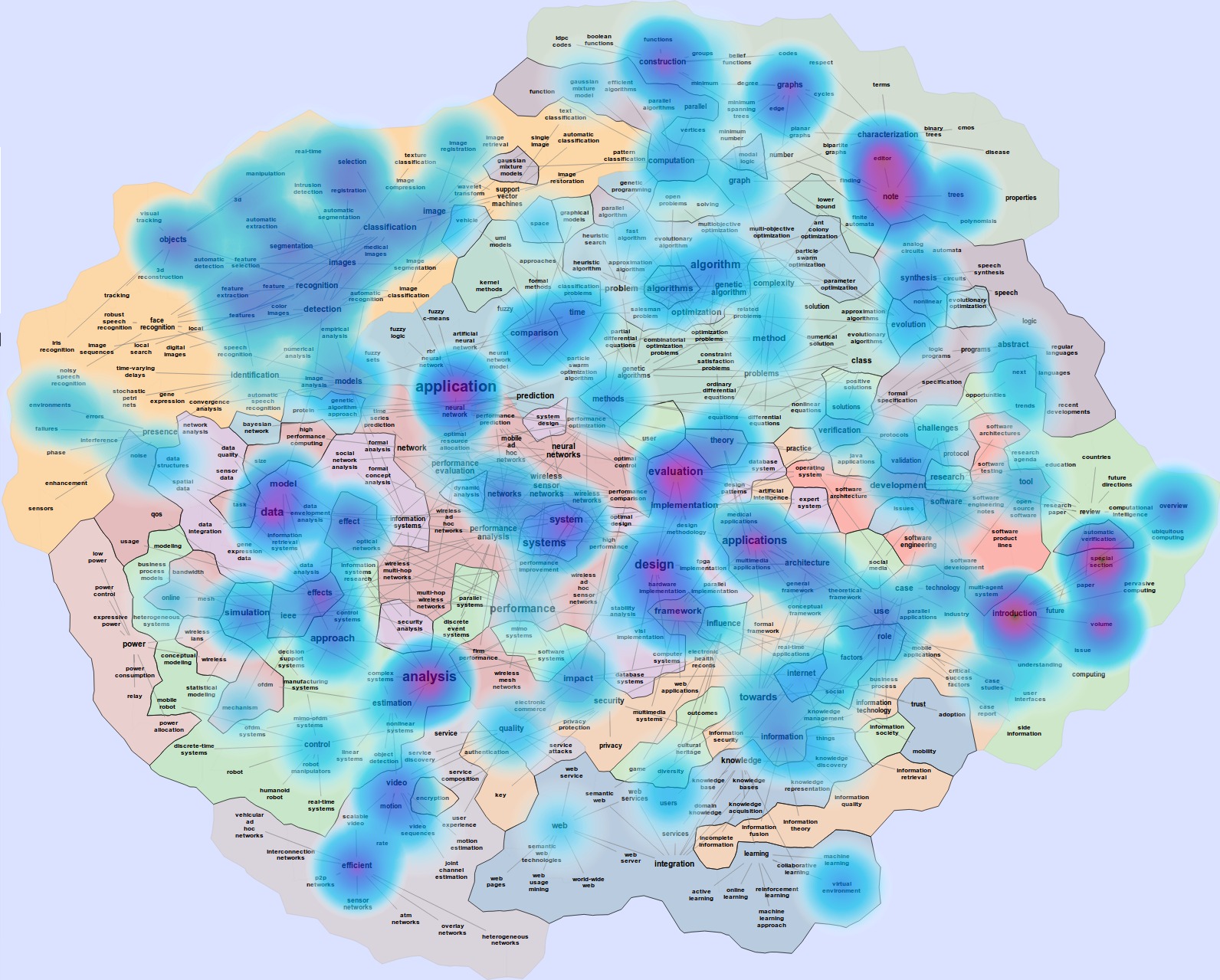}
        \label{tvcg_heatmap}
    }
    \caption{Conference and journal heatmaps overlaid on a map generated from
        70,000 paper titles, sampled uniformly from all available DBLP papers. Map
        generation algorithms are Multi-Word terms for term extraction, C-value
        With Unigrams for term ranking, Partial Match Jaccard Coefficient for similarity, and
        Pull Lesser Terms for filtering, with the number of top terms parameter set
    to 1500.}
\label{conf_journal_heatmaps}
\end{figure*}

The bibliographic information stored in DBLP allows us to plot heatmaps of
specific conferences and journals over a basemap of all documents. Fig.~\ref{conf_journal_heatmaps} shows heatmaps of papers from four venues: the
Computer Vision and Pattern Recognition conference (CVPR), the Symposium on
Theory of Computing (STOC), the International Conference on Web Services
(ICWS), and Transactions on Visualization and Computer Graphics (TVCG). These heatmaps are plotted from all available paper titles in
the DBLP database for each venue. The basemap over which the heatmaps are
plotted is made from 70,000 paper titles sampled uniformly from all entries in
DBLP. Some similarities can be seen between the venues: all share relatively high intensity in their heatmaps over terms ``application'', ``analysis'', ``method'', and ``evaluation''   Some notable topical
differences between venues also stand out. CVPR has a high intensity region in
the northwest corner of the map over terms such as ``images'', ``objects'', and
``recognition'', while STOC has most high intensity in the northeast corner of the
map, over terms related to ``graphs'', ``complexity'', and ``graphs''. ICWS has a high intensity in the south of the map over terms ``web services'' and ``systems'' while TVCG is literally all over the map, as visualization is associated with all areas of computing: from visualization of algorithms to algorithms for visualization, from design and analysis to applications and systems.

Effective heatmap coverage is a function both of the number of available
documents being plotted, and how well terms in the heatmap query set are
represented in the base map. Comparing the TVCG and ICWS heatmaps to the CVPR
and STOC heatmaps demonstrates this relationship between document availability
and heatmap coverage. Fewer papers are available for TVCG and ICWS in DBLP, causing
these venues have lesser representation in the basemap (which is constructed
from documents randomly sampled from all documents in DBLP), and so their
heatmaps cover less area. 

\subsection{Temporal Heatmap Overlays}
\begin{figure*}
    \centering
    \subfloat[Heatmap for 1954-1963 made from 399 paper titles]{
        \includegraphics[width=0.45\linewidth]{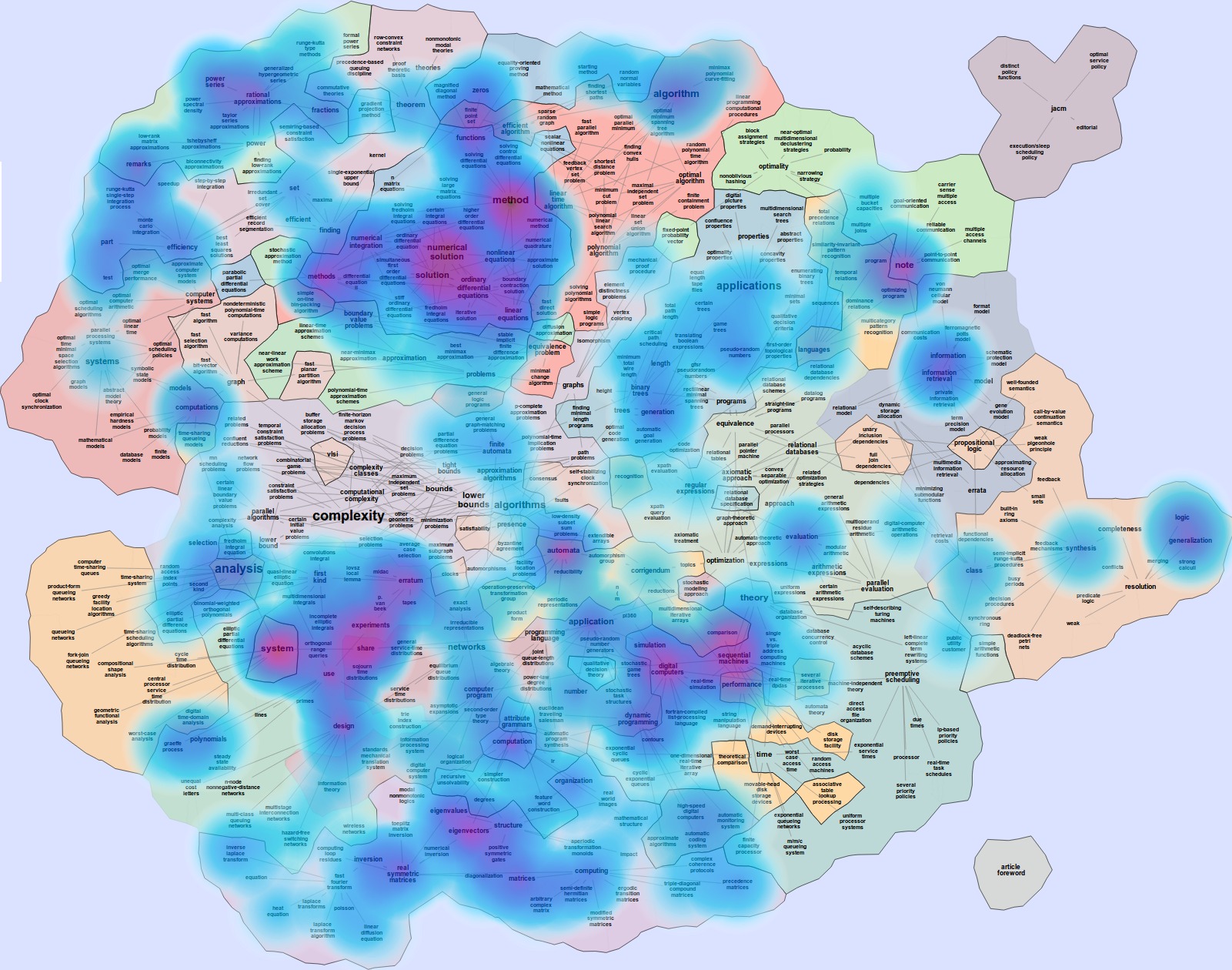}
        \label{}
    }
    \subfloat[Heatmap for 1964-1973 made from 400 paper titles]{
        \includegraphics[width=0.45\linewidth]{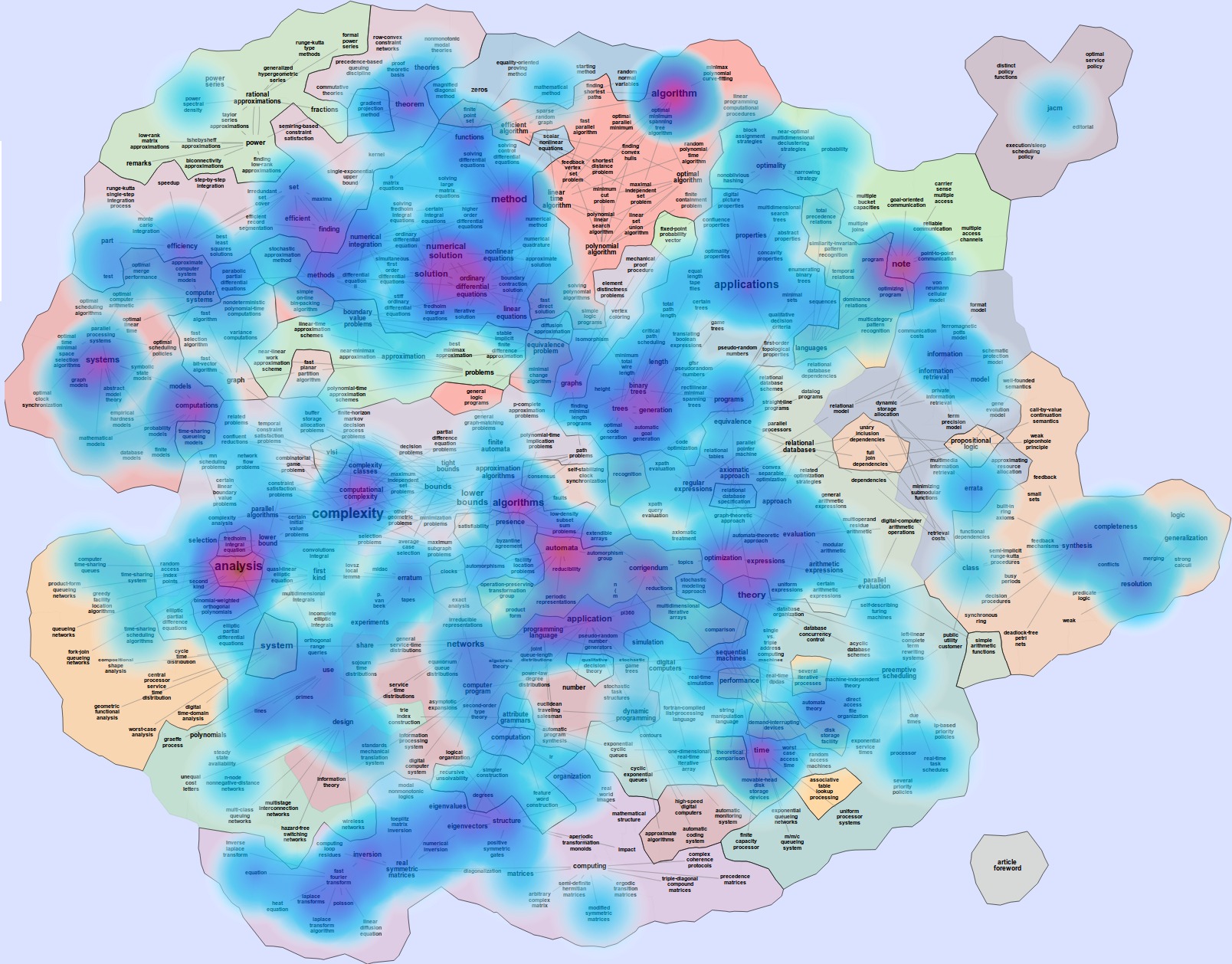}
        \label{}
    }

    \subfloat[Heatmap for 1974-1983 made from 400 paper titles]{
        \includegraphics[width=0.45\linewidth]{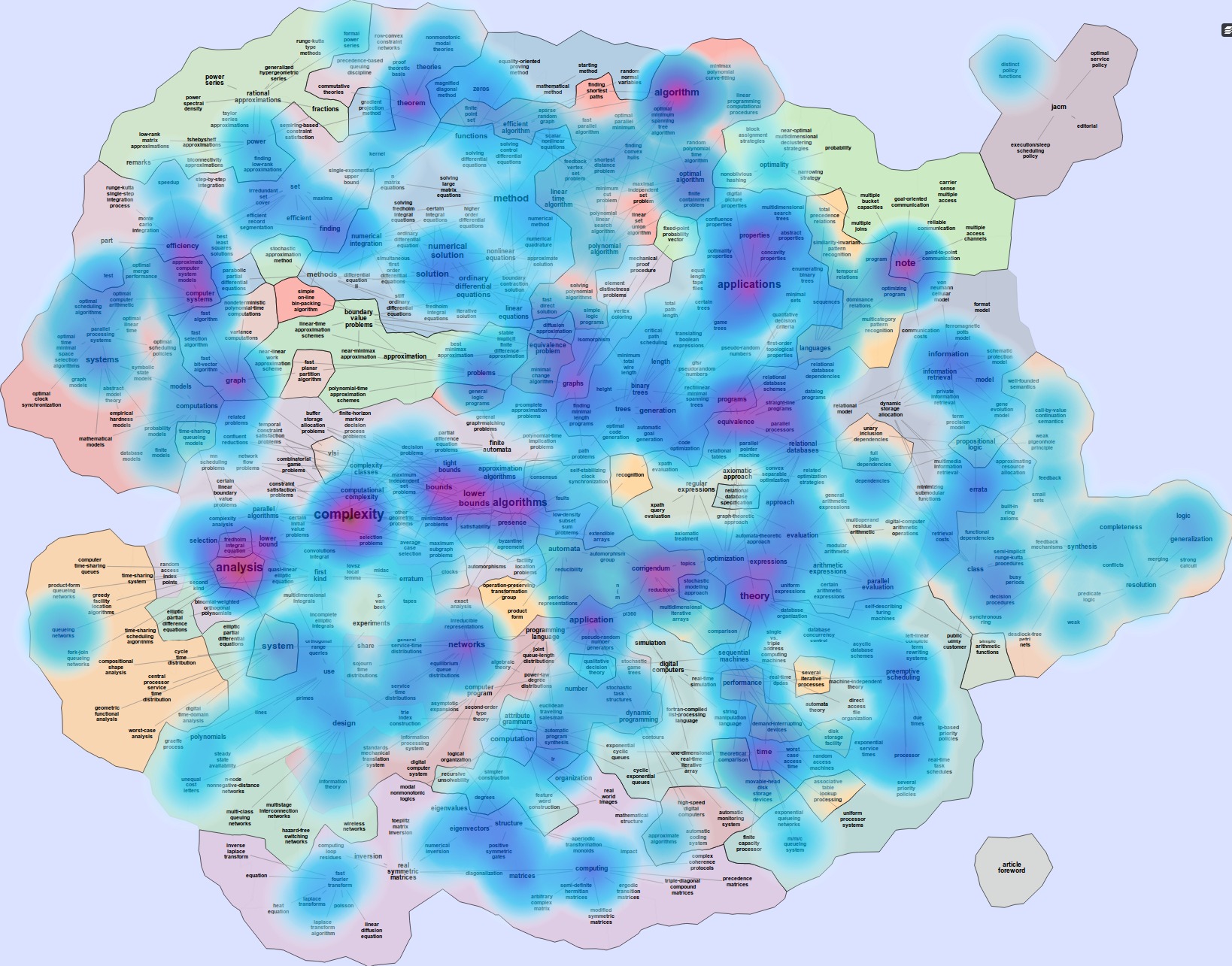}
        \label{}
    }
    \subfloat[Heatmap for 1984-1993 made from 400 paper titles]{
        \includegraphics[width=0.45\linewidth]{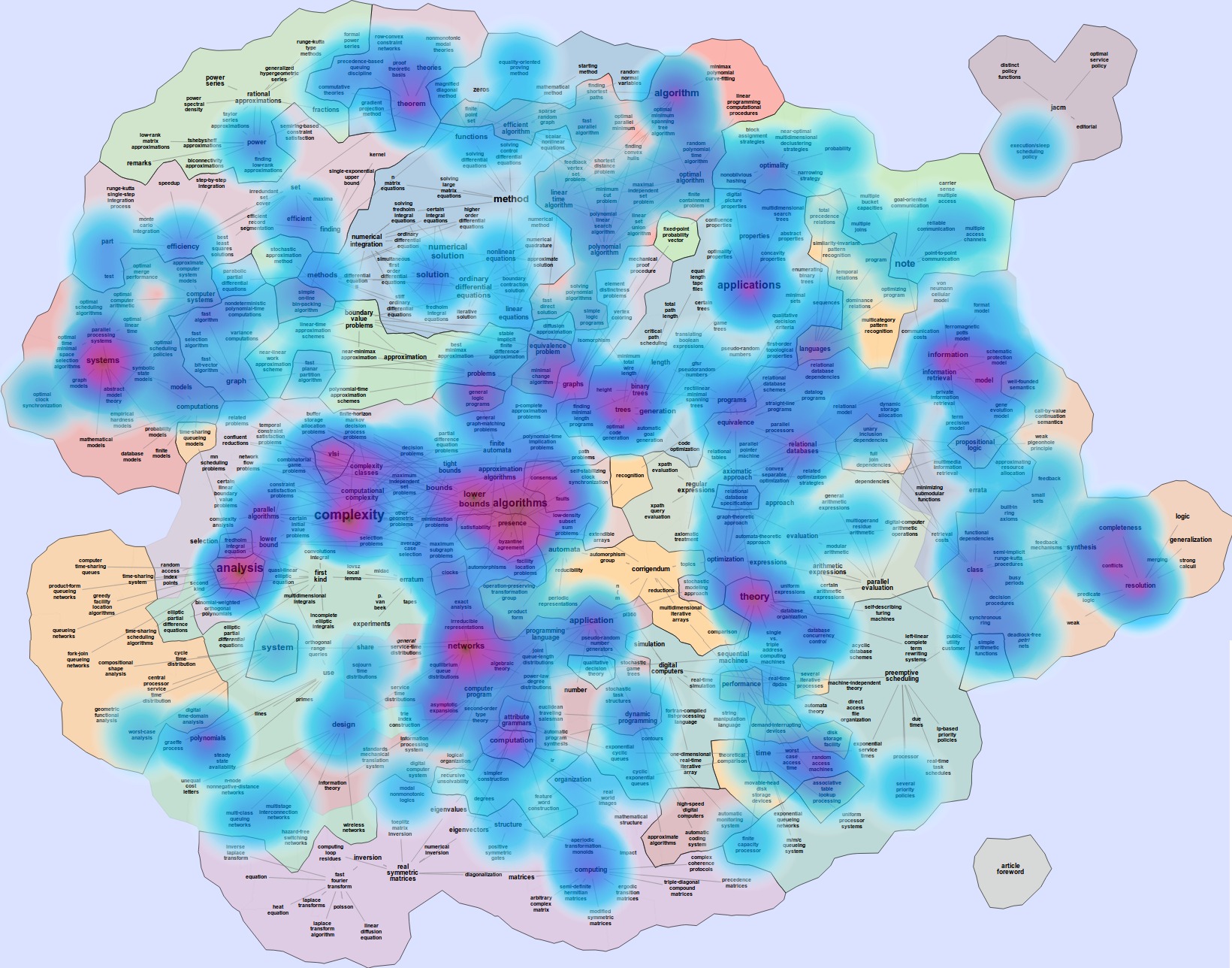}
        \label{}
    }

    \subfloat[Heatmap for 1994-2003 made from 372 paper titles]{
        \includegraphics[width=0.45\linewidth]{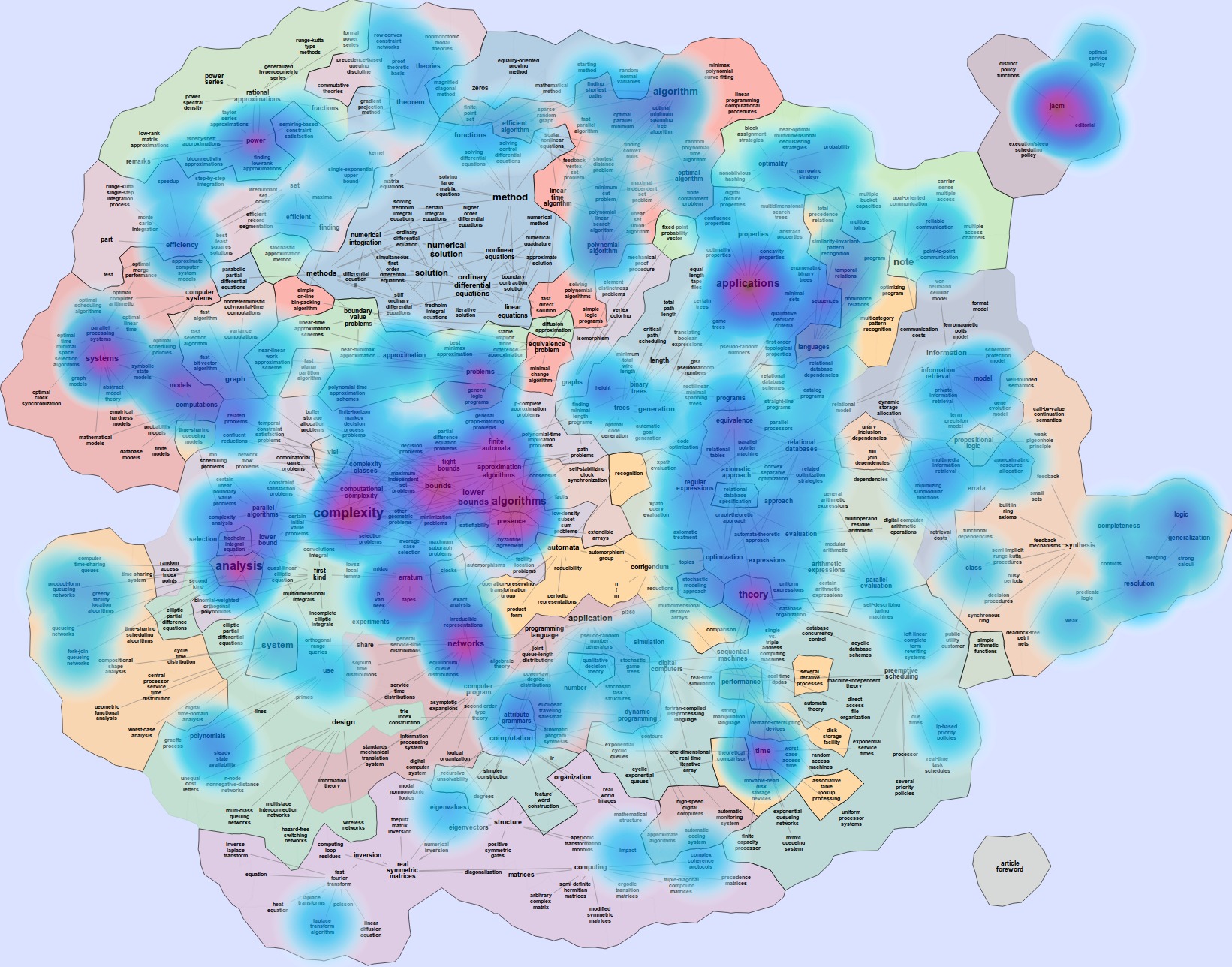}
        \label{}
    }
    \subfloat[Heatmap for 2004-2013 made from 284 paper titles]{
        \includegraphics[width=0.45\linewidth]{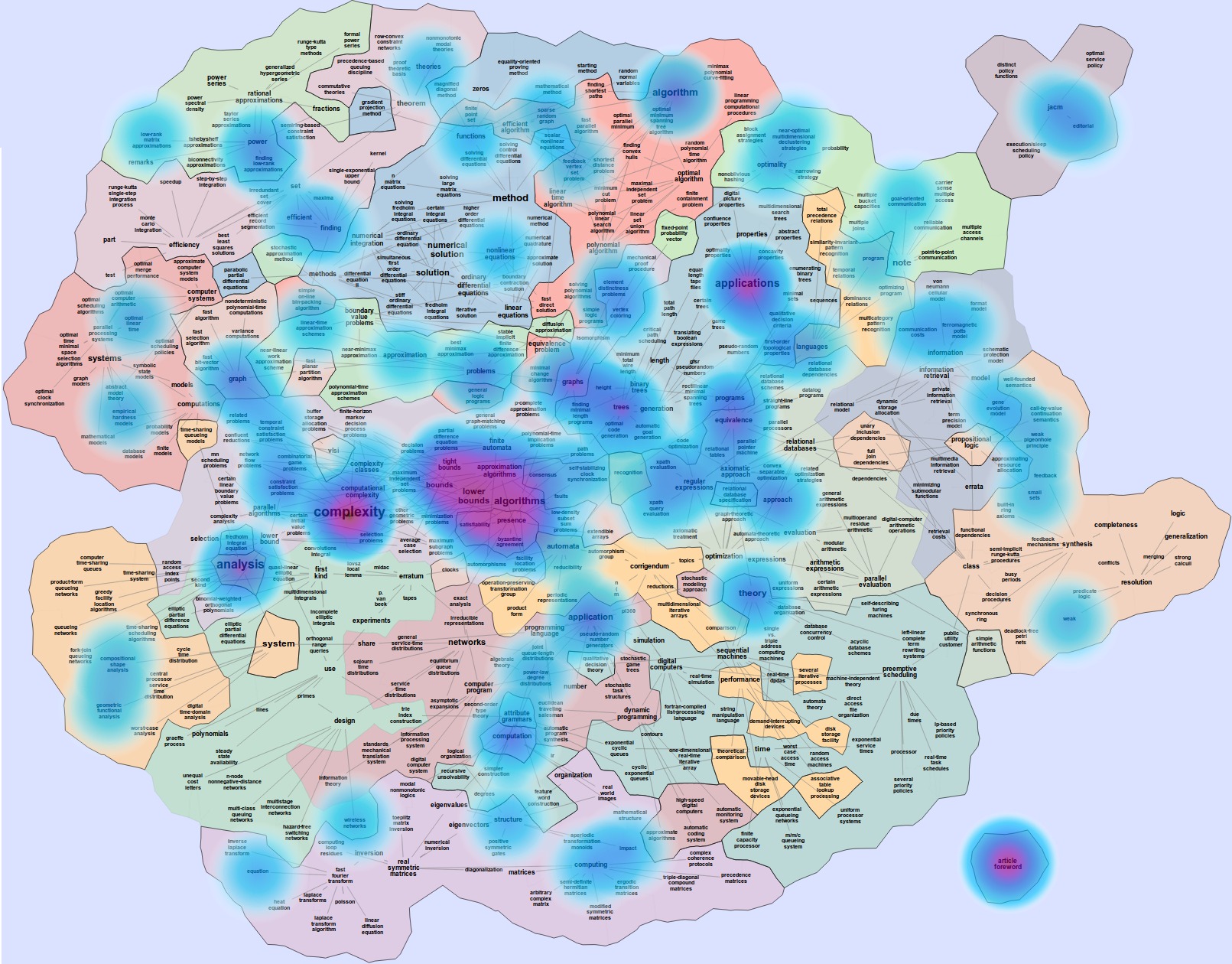}
        \label{}
    }
\caption{Heatmaps of six decades of papers from Journal of the ACM (JACM).
    Basemap is generated from multi-word terms extracted from the titles of
    1,998 paper titles published in JACM, using the C-Value with Unigrams ranking,
    Partial Match Jaccard Coefficient similarity, and Pull Lesser Terms filtering
    functions. A maximum of 400 paper titles were sampled from the JACM's publications
    for each decade.}
\label{jacm_temporal_heatmaps}
\end{figure*}

Specifying different date ranges for heatmap queries allows the generation of
maps that show how areas of research have spread across the topic basemaps over
time. The maps in Fig.~\ref{jacm_temporal_heatmaps} show how terms in the
titles of papers published in the Journal of the ACM (JACM) have shifted over
the past six decades, starting in 1954. The heatmap for papers from 1954-1963
has high intensity values over terms dealing with numerical and matrix methods.
Computational complexity grows in intensity in the 1964-1973 map, and complexity
and algorithmic bounds outpace numerical methods in 1974-1983. The algorithmic
bound terms remain consistently intense throughout the remaining decades. An
easy to notice trend is that the focus of the journal has noticeably narrowed
over time: in the first four decades the topics are all over the map, but in
the last decade the topics are concentrated around complexity, algorithms, and
bounds.

\subsection{Individual Paper Heatmaps}
\begin{figure}
  \includegraphics[width=0.9\linewidth]{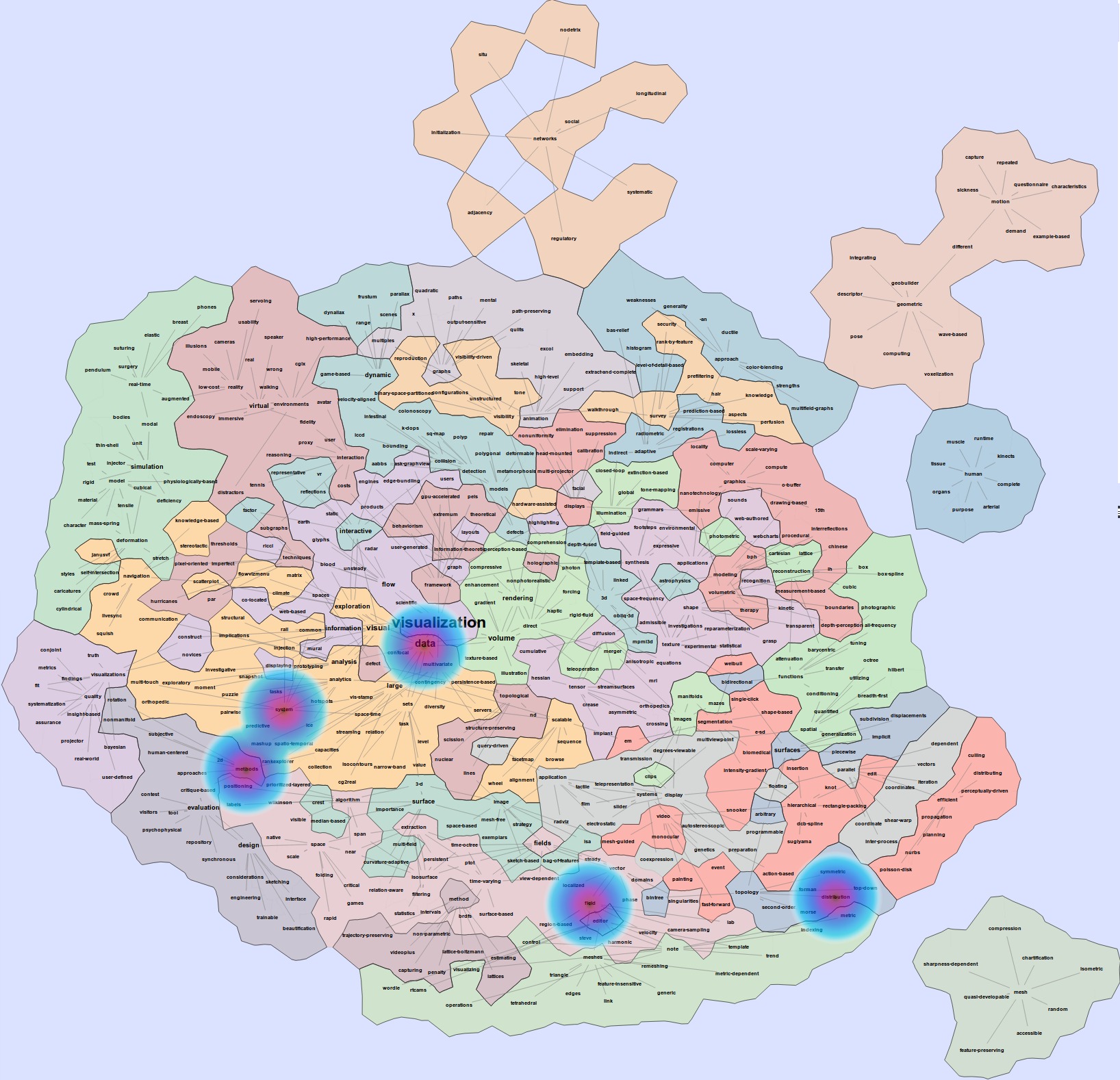}
  \caption{A heatmap from the abstract of this paper,
      over a basemap from 1,343 TVCG paper titles, using the TF
      ranking, LSA similarity, and Pull Lesser Terms filtering functions, with
  the number of terms parameter set to 1700.}
  \label{abstract_heatmap}
\end{figure}

To construct a heatmap visualization of the topics in a single paper, we can
run the same term extraction algorithms outlined above on the abstract or full
body text of a paper. This heatmap is then overlaid on a basemap constructed
from DBLP paper titles, as above. Figure~\ref{abstract_heatmap} shows a heatmap
constructed from terms in the abstract of this paper, over a single-word basemap
of TVCG paper titles.

\section{Implementation}
\subsection{Modularity}
The system is built with a modular design to accommodate future incorporation
of additional natural language processing algorithms.  Each of the stages of
the map generation pipeline (ranking, similarity, and filtering) is handled by
a separate module of code. Within each module, the functions that perform the
module's task are designed to be substitutable, taking standardized input and
output. We plan to expand the system's capabilities by testing the ability of
other ranking, similarity, and filtering algorithms to produce maps that
provide a better visual representation of the underlying topic space.  Source
code for the system is available for others who wish to experiment with
algorithms of their own.

\subsection{Database}
Paper titles and meta-information are stored in a SQL database, containing
entries for 2,184,270 papers, journal articles, conference proceedings, theses,
and books. This bibliographic information is parsed from an XML dump of DBLP
entries, containing author, conference or journal, and date meta-information
for each paper title~\cite{ley2009dblp}.  There are over one million personal
web pages listed in DBLP, with title ``Home Page'', and tag information is used
to exclude these. Additionally, as DBLP contains papers with titles in several
languages, an effort is made to detect the language of each title, using a
trigram character classifier. Titles classified as English by this classifier
are marked in the database, and only these titles are currently used in the map
generation. Each paper is associated with its author and  journal or conference
if this information is available in DBLP. The database contains records for
1,324 journals, 6,904 conferences, and 1,237,445 authors which can be used to
filter document title queries for map construction.

\subsection{Server}
The system is implemented using Python 2.7. Full source code (for map making,
the DBLP database interface, and the web server) is available at
\url{github.com/dpfried/mocs}. Natural language
processing code for term extraction is implemented using utilities from the
NLTK~\cite{bird2009natural} library. The NumPy and SciPy~\cite{scipy} numerical
computation libraries are used for implementing the similarity functions and
ranking algorithms. The server is hosted in Django, using Celery as a back-end
task manager, and SQLAlchemy for database interface. Maps are displayed in the
user's browser using SVG rendering capabilities of AT\&T's GraphViz system~\cite{graphviz}.  These SVG elements
are rendered in a zoomable and pannable container provided by the open source
OpenLayers JavaScript display library~\cite{openlayers}.  Heatmaps are overlaid with the heatmap plugin in OpenLayers, together with additional JavaScript that calculates term positions and SVG coordinate transforms, in order to correctly position the heatmap over the basemap when zooming.

\section{Conclusions and Future Work\label{sec_conc}}

In this paper we presented a practical approach for visualizing large-scale a
bibliographic data via natural language processing and using a geographic map
metaphor.  We described the {\tt MoCS} system in the context of the DBLP
bibliography server and demonstrated several possible explorative visualization
uses of the system. The novel aspects of the system include modifications to
natural language processing techniques (allowing us to work with only titles of
research papers), the ability to combine arbitrary basemaps with heatmap
overlays (showing temporal evolution, or profiles of conferences and journals),
and the modularity and availability of the interactive visualization system
(making it possible to experiment with different approaches to various
subproblems). An interactive interface to the system, and a video of the system
in action, are available at \url{mocs.cs.arizona.edu}.

There are likely more possible uses of such a visualization system. For
example, many journals (e.g., Cell, Earth and Planetary Science, Molecular
Phylogenomics and Evolution) have recently added requirements for graphical
abstracts as a part of research papers. These are single-panel images designed
to give readers an immediate understanding of the take-home message of the
paper. {\tt MoCS} can be used to generate graphical abstracts using a basemap
from the journal and heatmap of the submission.

We would have liked to compare the performance of our system against earlier
and related approaches. However, this is nearly impossible as very few such
systems are fully functional online or provides source code. We contacted the
authors of a dozen earlier semantic word-cloud or spatialization based systems
but none were able to share source code or executables. 

While ours is indeed a functional system, and it does offer various options for
the natural language processing step, for the generation of the graph, and for
the final map rendering, there are many possible future directions:
\begin{enumerate}

\item We would like to experimentally verify whether our maps based on research paper titles correspond to what experts in the field expect to see. If not, topic models for term extraction and similarity that incorporate lexical priors, or words that are of specific interest can be used~\cite{jagarlamudiDU12}. Thus, we could specify ``seed words'' or ``start words'' that must included in the map. 

\item How much additional information
  and precision can be gained from abstracts compared to just titles
  of papers? Similarly, what is the additional information gain when
  going from abstracts to entire papers?

\item We can study departmental, state-wide, and even country-wide profiles over the base map of CS. This would hopefully allow us to visually compare and contrast the type of research done in different universities, states, and countries. 

\item Automatically labeling countries on the map could be accomplished by looking for the most frequent conferences and journals with topics in a particular country, and extracting the top 2-3 relevant terms.

\item Statistical methods for multi-word term extraction and ranking, such as     topical $n$-grams~\cite{wang2007topical} or LexRank~\cite{erkan2004lexrank} may allow us to produce terms that are more representative of topics in the document titles than the terms extracted through POS tagging and pattern matching alone.

\item Only terms that appear in both the basemap and heatmap queries are    currently displayed in heatmaps. To create heatmaps that also cover related   terms in the map, the pairwise term similarity values could be used to    diffuse heatmap intensity onto terms that were unseen in the heatmap query, but that are similar to those seen in the query.

\item The graph embedding and graph clustering combinations that are available in GMap often result in fragmented maps. We would like to expand  the functionality of GMap by providing cluster-based (and thus non-fragmented) embedding.

\item The methodology described here is not limited to computer
  science research papers. It should be possible to generalize to other research
  areas, starting with physics (due to ArXiv) and medicine (PubMed).
\end{enumerate}

\acknowledgements{We thank Henry Kerschen for help with the the webpage for
    {\em maps of computer science} server: \url{mocs.cs.arizona.edu}. We also
    thank Stephan Diehl, Sandiway Fong, Yifan Hu, and David Sidi for discussions about this
    project and ongoing system evaluation.}
\bibliographystyle{abbrv}
\bibliography{refs}

\begin{thebibliography}{10}

\bibitem{openlayers}
{OpenLayers}: Free maps for the web.
\newblock http://www.openlayers.org/.

\bibitem{infosky}
K.~Andrews, W.~Kienreich, V.~Sabol, J.~Becker, G.~Droschl, F.~Kappe,
  M.~Granitzer, P.~Auer, and K.~Tochtermann.
\newblock The infosky visual explorer: exploiting hierarchical structure and
  document similarities.
\newblock {\em Information Visualization}, 1(3-4):166--181, 2002.

\bibitem{bird2009natural}
S.~Bird, E.~Klein, and E.~Loper.
\newblock {\em Natural language processing with Python}.
\newblock O'Reilly Media, Incorporated, 2009.

\bibitem{Blei03}
D.~M. Blei, A.~Y. Ng, and M.~I. Jordan.
\newblock Latent dirichlet allocation.
\newblock {\em Journal of Machine Learning Research}, 3:993--1022, 2003.

\bibitem{boyd2010syntactic}
J.~Boyd-Graber and D.~M. Blei.
\newblock Syntactic topic models.
\newblock {\em arXiv preprint arXiv:1002.4665}, 2010.

\bibitem{Byelas_2006_map}
H.~Byelas and A.~Telea.
\newblock Visualization of areas of interest in software architecture diagrams.
\newblock In {\em ACM Symp.~on Software Visualization (SoftVis'06)}, pages
  105--114, 2006.

\bibitem{facetatlas}
N.~Cao, J.~Sun, Y.-R. Lin, D.~Gotz, S.~Liu, and H.~Qu.
\newblock Facetatlas: Multifaceted visualization for rich text corpora.
\newblock {\em IEEE Transactions on Visualization and Computer Graphics},
  16(6):1172--1181, 2010.

\bibitem{Collins_2009_bubblset}
C.~Collins, G.~Penn, and S.~Carpendale.
\newblock Bubble sets: Revealing set relations with isocontours over existing
  visualizations.
\newblock {\em IEEE TVCG}, 15(6):1009--1016, 2009.

\bibitem{collins-09}
C.~Collins, F.~B. Vi{\'e}gas, and M.~Wattenberg.
\newblock Parallel tag clouds to explore and analyze faceted text corpora.
\newblock In {\em IEEE VAST}, pages 91--98, 2009.

\bibitem{Cortese_2006_contour}
P.~F. Cortese, G.~D. Battista, A.~Moneta, M.~Patrignani, and M.~Pizzonia.
\newblock Topographic visualization of prefix propagation in the internet.
\newblock {\em IEEE TVCG}, 12:725--732, 2006.

\bibitem{Cui_2010_wordcloud}
W.~Cui, Y.~Wu, S.~Liu, F.~Wei, M.~X. Zhou, and H.~Qu.
\newblock Context-preserving, dynamic word cloud visualization.
\newblock {\em Computer Graphics and Applications}, 30:42--53, 2010.

\bibitem{deerwester1990indexing}
S.~Deerwester, S.~T. Dumais, G.~W. Furnas, T.~K. Landauer, and R.~Harshman.
\newblock Indexing by latent semantic analysis.
\newblock {\em Journal of the American society for information science},
  41(6):391--407, 1990.

\bibitem{graphviz}
J.~Ellson, E.~R. Gansner, E.~Koutsofios, S.~C. North, and G.~Woodhull.
\newblock Graphviz and dynagraph—static and dynamic graph drawing tools.
\newblock In {\em Graph Drawing Software}, pages 127--148. Springer, 2004.

\bibitem{erkan2004lexrank}
G.~Erkan and D.~R. Radev.
\newblock Lexrank: Graph-based lexical centrality as salience in text
  summarization.
\newblock {\em J. Artif. Intell. Res. (JAIR)}, 22:457--479, 2004.

\bibitem{Fabrikant_2006_map_infoviz}
S.~I. Fabrikant, D.~R. Montello, and D.~M. Mark.
\newblock The distance-similarity metaphor in region-display spatializations.
\newblock {\em IEEE Computer Graphics \& Application}, 26:34--44, 2006.

\bibitem{forbesinteractive}
A.~Forbes, B.~Alper, T.~H{\"o}llerer, and G.~Legrady.
\newblock Interactive folksonomic analytics with the tag river visualization.
\newblock In {\em Interactive Visual Text Analytics for Decision Making
  Workshop}. 2011.

\bibitem{francis1979manual}
W.~N. Francis and H.~Ku{\v{c}}era.
\newblock {\em Manual of information to accompany a standard corpus of
  present-day edited American English, for use with digital computers}.
\newblock Brown University, Department of Lingustics, 1979.

\bibitem{frantzi2000automatic}
K.~Frantzi, S.~Ananiadou, and H.~Mima.
\newblock Automatic recognition of multi-word terms:. the c-value/nc-value
  method.
\newblock {\em International Journal on Digital Libraries}, 3(2):115--130,
  2000.

\bibitem{Fruchterman_Reingold_1991}
T.~Fruchterman and E.~Reingold.
\newblock Graph drawing by force directed placement.
\newblock {\em Software-Practice and Experience}, 21:1129--1164, 1991.

\bibitem{Emden_Hu_recsys_2009}
E.~Gansner, Y.~Hu, S.~Kobourov, and C.~Volinsky.
\newblock Putting recommendations on the map - visualizing clusters and
  relations.
\newblock In {\em 3rd ACM Conf.~on Recommender Systems}, pages 345--348, 2009.

\bibitem{topicnets}
B.~Gretarsson, J.~O'Donovan, S.~Bostandjiev, T.~H{\"o}llerer, A.~U. Asuncion,
  D.~Newman, and P.~Smyth.
\newblock Topicnets: Visual analysis of large text corpora with topic modeling.
\newblock {\em ACM TIST}, 3(2):23, 2012.

\bibitem{harrower}
M.~Harrower.
\newblock Tips for designing effective animated maps.
\newblock {\em Cartographic Perspectives}, 44:63--65, 2003.

\bibitem{themeriver}
S.~Havre, E.~G. Hetzler, P.~Whitney, and L.~T. Nowell.
\newblock Themeriver: Visualizing thematic changes in large document
  collections.
\newblock {\em IEEE Trans. Vis. Comput. Graph.}, 8(1):9--20, 2002.

\bibitem{hu99visualizing}
Y.~Hu, E.~Gansner, and S.~Kobourov.
\newblock {Visualizing Graphs and Clusters as Maps}.
\newblock {\em IEEE Computer Graphics and Applications}, 99(1):54--66, 2010.

\bibitem{jaccard1901etude}
P.~Jaccard.
\newblock {\em Etude comparative de la distribution florale dans une portion
  des Alpes et du Jura}.
\newblock Impr. Corbaz, 1901.

\bibitem{jagarlamudiDU12}
J.~Jagarlamudi, H.~{Daum{\'e} III}, and R.~Udupa.
\newblock Incorporating lexical priors into topic models.
\newblock In {\em EACL}, pages 204--213, 2012.

\bibitem{scipy}
E.~Jones, T.~Oliphant, P.~Peterson, et~al.
\newblock {SciPy}: Open source scientific tools for {Python}, 2001--.

\bibitem{maniwordle}
K.~Koh, B.~Lee, B.~H. Kim, and J.~Seo.
\newblock Maniwordle: Providing flexible control over wordle.
\newblock {\em IEEE Trans. Vis. Comput. Graph.}, 16(6):1190--1197, 2010.

\bibitem{mds}
J.~B. Kruskal and M.~Wish.
\newblock {\em Multidimensional Scaling}.
\newblock Sage Press, 1978.

\bibitem{HKK96}
K.~Lagus, T.~Honkela, S.~Kaski, and T.~Kohonen.
\newblock Self-organizing maps of document collections: A new approach to
  interactive exploration.
\newblock In {\em KDD}, pages 238--243, 1996.

\bibitem{ley2009dblp}
M.~Ley.
\newblock {DBLP} - some lessons learned.
\newblock {\em PVLDB}, 2(2):1493--1500, 2009.

\bibitem{tiara}
S.~Liu, M.~X. Zhou, S.~Pan, W.~Qian, W.~Cai, and X.~Lian.
\newblock Interactive, topic-based visual text summarization and analysis.
\newblock In {\em Proceedings of the 18th ACM conference on Information and
  knowledge management}, pages 543--552, 2009.

\bibitem{manning2008introduction}
C.~D. Manning, P.~Raghavan, and H.~Sch{\"u}tze.
\newblock {\em Introduction to information retrieval}, volume~1.
\newblock Cambridge University Press Cambridge, 2008.

\bibitem{topicislands}
N.~E. Miller, P.~C. Wong, M.~Brewster, and H.~Foote.
\newblock Topic islands - a wavelet-based text visualization system.
\newblock In {\em IEEE Visualization}, pages 189--196, 1998.

\bibitem{Newman_2006}
M.~E.~J. Newman.
\newblock Modularity and community structure in networks.
\newblock {\em Proc. Natl. Acad. Sci. USA}, 103:8577--8582, 2006.

\bibitem{brandes12}
A.~Nocaj and U.~Brandes.
\newblock Organizing search results with a reference map.
\newblock {\em {IEEE} Transactions on Visualization and Computer Graphics},
  18(12):2546--2555, 2012.

\bibitem{CGF:CGF3107}
F.~V. Paulovich, F.~M.~B. Toledo, G.~P. Telles, R.~Minghim, and L.~G. Nonato.
\newblock Semantic wordification of document collections.
\newblock {\em Computer Graphics Forum}, 31(3):1145--1153, 2012.

\bibitem{clouds-07}
A.~W. Rivadeneira, D.~M. Gruen, M.~J. Muller, and D.~R. Millen.
\newblock Getting our head in the clouds: toward evaluation studies of
  tagclouds.
\newblock In {\em CHI}, pages 995--998, 2007.

\bibitem{treemap}
B.~Shneiderman and M.~Wattenberg.
\newblock Ordered treemap layouts.
\newblock In {\em INFOVIS}, pages 73--78, 2001.

\bibitem{Simonetto_2009_overlapset}
P.~Simonetto, D.~Auber, and D.~Archambault.
\newblock Fully automatic visualisation of overlapping sets.
\newblock {\em Computer Graphics Forum}, 28:967--974, 2009.

\bibitem{sf-sm-03}
A.~Skupin and S.~I. Fabrikant.
\newblock Spatialization methods: a cartographic research agenda for
  non-geographic information visualization.
\newblock {\em Cartography and Geographic Information Science}, 30:95--119,
  2003.

\bibitem{phrasenet}
F.~van Ham, M.~Wattenberg, and F.~B. Vi{\'e}gas.
\newblock Mapping text with phrase nets.
\newblock {\em IEEE Trans. Vis. Comput. Graph.}, 15(6):1169--1176, 2009.

\bibitem{viegas-tag-08}
F.~B. Vi{\'e}gas and M.~Wattenberg.
\newblock Timelines - tag clouds and the case for vernacular visualization.
\newblock {\em Interactions}, 15(4):49--52, 2008.

\bibitem{wordle}
F.~B. Vi{\'e}gas, M.~Wattenberg, and J.~Feinberg.
\newblock Participatory visualization with wordle.
\newblock {\em IEEE Trans. Vis. Comput. Graph.}, 15(6):1137--1144, 2009.

\bibitem{CGF:CGF1898}
T.~von Landesberger, A.~Kuijper, T.~Schreck, J.~Kohlhammer, J.~van Wijk, J.-D.
  Fekete, and D.~Fellner.
\newblock Visual analysis of large graphs: State-of-the-art and future research
  challenges.
\newblock {\em Computer Graphics Forum}, 30(6):1719--1749, 2011.

\bibitem{wang2007topical}
X.~Wang, A.~McCallum, and X.~Wei.
\newblock Topical n-grams: Phrase and topic discovery, with an application to
  information retrieval.
\newblock In {\em Data Mining, 2007. ICDM 2007. Seventh IEEE International
  Conference on}, pages 697--702. IEEE, 2007.

\bibitem{857579}
J.~A. Wise, J.~J. Thomas, K.~Pennock, D.~Lantrip, M.~Pottier, A.~Schur, and
  V.~Crow.
\newblock Visualizing the non-visual: spatial analysis and interaction with
  information from text documents.
\newblock In {\em IEEE Symp. on Information Visualization}, pages 51--58, 1995.

\bibitem{wu2011semantic}
Y.~Wu, T.~Provan, F.~Wei, S.~Liu, and K.-L. Ma.
\newblock Semantic-preserving word clouds by seam carving.
\newblock In {\em Computer Graphics Forum}, volume~30, pages 741--750, 2011.

\bibitem{zamir1999grouper}
O.~Zamir and O.~Etzioni.
\newblock Grouper: a dynamic clustering interface to web search results.
\newblock {\em Computer Networks}, 31(11):1361--1374, 1999.

\end{thebibliography}
\end{document}